\documentclass[superscriptaddress,english,floatfix,twocolumn,showpacs,aps,prb,reprint]{revtex4-2}

\usepackage{amsmath}
\usepackage{graphicx}
\usepackage{bm}
\usepackage{dsfont}
\usepackage{xfrac}
\usepackage{siunitx}
\usepackage{nicefrac} 
\usepackage{hyperref}
\usepackage[normalem]{ulem}
\usepackage{braket}
\usepackage{booktabs}

\usepackage{tabularx,ragged2e}
\usepackage{xcolor,colortbl}

\usepackage{xcolor} 

\hypersetup{colorlinks=true, citecolor=blue, urlcolor=blue, linkcolor=blue, breaklinks=true}

\definecolor{forestgreen}{rgb}{0.13, 0.55, 0.13}

\begin{document}

\title{Competing magnetic fluctuations and orders in a multiorbital model of doped \texorpdfstring{SrCo$_2$As$_2$}{}}

\author{Ana-Marija Nedi\' c}
\affiliation{Department of Physics and Astronomy, Iowa State University, Ames, Iowa 50011, USA}
\affiliation{Ames National Laboratory, Ames, Iowa 50011, USA}

\author{Morten H. Christensen}
\affiliation{School of Physics and Astronomy, University of Minnesota, Minneapolis, MN 55455, USA}
\affiliation{Niels Bohr Institute, University of Copenhagen, Denmark}

\author{Y.~Lee}
\affiliation{Department of Physics and Astronomy, Iowa State University, Ames, Iowa 50011, USA}
\affiliation{Ames National Laboratory, Ames, Iowa 50011, USA}

\author{Bing Li}
\affiliation{Department of Physics and Astronomy, Iowa State University, Ames, Iowa 50011, USA}
\affiliation{Ames National Laboratory, Ames, Iowa 50011, USA}

\author{Benjamin~G.~Ueland}
\affiliation{Department of Physics and Astronomy, Iowa State University, Ames, Iowa 50011, USA}
\affiliation{Ames National Laboratory, Ames, Iowa 50011, USA}

\author{Rafael~M.~Fernandes}
\affiliation{School of Physics and Astronomy, University of Minnesota, Minneapolis, MN 55455, USA}

\author{Robert~J.~McQueeney }
\affiliation{Department of Physics and Astronomy, Iowa State University, Ames, Iowa 50011, USA}
\affiliation{Ames National Laboratory, Ames, Iowa 50011, USA}

\author{Liqin~Ke}
\affiliation{Ames National Laboratory, Ames, Iowa 50011, USA}

\author{Peter~P.~Orth}
\email{porth@iastate.edu}
\affiliation{Department of Physics and Astronomy, Iowa State University, Ames, Iowa 50011, USA}
\affiliation{Ames National Laboratory, Ames, Iowa 50011, USA}

\date{\today}

\begin{abstract}
We revisit the intriguing magnetic behavior of the paradigmatic itinerant frustrated magnet $\rm{Sr}\rm{Co}_2\rm{As}_2$, which shows strong and competing magnetic fluctuations yet does not develop long-range magnetic order.
By calculating the static spin susceptibility $\chi(\mathbf{q})$ within a realistic sixteen orbital Hubbard-Hund model, we determine the leading instability to be ferromagnetic (FM). We then explore the effect of doping and calculate the critical Hubbard interaction strength $U_c$ that is required for the development of magnetic order. We find that $U_c$ decreases under electron doping and with increasing Hund's coupling $J$, but increases rapidly under hole doping. This suggests that magnetic order could possibly emerge under electron doping but not under hole doping, which agrees with experimental findings. We map out the leading magnetic instability as a function of doping and Hund's coupling and find several antiferromagnetic phases in addition to FM. We also quantify the degree of itinerant frustration in the model and resolve the contributions of different orbitals to the magnetic susceptibility. Finally, we discuss the dynamic spin susceptibility, $\chi(\mathbf{q}, \omega)$, at finite frequencies, where we recover the anisotropy of the peaks at $\mathbf{Q}_\pi = (\pi, 0)$ and $(0, \pi)$ observed by inelastic neutron scattering that is associated with the phenomenon of itinerant magnetic frustration. By comparing results between theory and experiment, we conclude that the essential experimental features of doped SrCo$_2$As$_2$ are well captured by an itinerant Hubbard-Hund multiorbital model if one considers a small shift of the chemical potential towards hole doping. 
\end{abstract}
\maketitle
\section{Introduction}
\label{sec:intro}
The tetragonal 122 cobalt arsenides SrCo$_2$As$_2$, CaCo$_{2-y}$As$_2$, and BaCo$_2$As$_2$ are members of a wider class of cobalt pnictides that exhibit strong competition between ferromagnetic (FM) and stripe-type antiferromagnetic (AF) fluctuations in the square Co layers. 
Unlike the structurally and chemically similar 122 iron-based systems AFe$_2$As$_2$ (A = Ca, Sr, Ba) that exhibit long-range stripe-type AF order in the Fe layers~\cite{Johnston2010}, the cobalt arsenides show either FM order in the Co layers, e.g. CaCo$_{2-y}$As$_2$~\cite{Cheng2012, Quirinale2013, Anand2014_1,Jayasekara2017}, or remain paramagnetic (PM) down to the lowest temperatures measured, e.g. SrCo$_2$As$_2$~\cite{Pandey2013, Wiecki-SrCo2As2-PRB-2015, Li2019_Sr} and BaCo$_2$As$_2$~\cite{Sefat2009,Anand2014_2}. Different stackings of the two-dimensional (2D) FM planes are observed. While CaCo$_{2-y}$As$_2$ shows AF stacking (A-type), one finds 3D FM in Sr$_{1-x}$La$_x$Co$_2$As$_2$~\cite{Shen2018} and a more complex helical stacking of the FM planes in Sr(Co$_{1-x}$Ni$_x$)$_2$As$_2$~\cite{Wilde2019} and Ca$_{1-x}$Sr$_x$Co$_2$As$_2$~\cite{Jayasekara2013, Ying_EPL_2013,Sangeetha_CaSrCo2As2_PRL_2017, Li2019_Ca}.

Stripe-type AF fluctuations are believed to play a key role for the emergence of superconductivity in the Fe-pnictides~\cite{paglionenatphys2010,chubukovannrevcondmatphys2012}. Inelastic neutron scattering (INS)~\cite{Sapkota2017, Yu_Dai_2019_Sr, Li2019_Sr} and NMR~\cite{Wiecki-SrCo2As2-PRB-2015} measurements have revealed that strong stripe-AF fluctuations are also present in the Co-arsenides. Contrary to the doped Fe-based materials, the Co-based materials, however, do not exhibit superconductivity under slight doping. This may be related to the coexistence of strong FM fluctuations and long-range FM order, which generally tend to inhibit superconducting singlet pairing. It is therefore interesting to explore the fate and possible suppression of FM with electronic doping. 

A general underlying open question is what causes the observed competition of stripe-AF and FM fluctuations in the cobalt arsenides, whose Fermi surfaces lack clear nesting wavevectors
~\cite{Pandey2013, Mao-PRB-2018, Li2019_Sr} 
In previous works, this competition was phenomenologically captured within a frustrated local-moment $J_1$-$J_2$ Heisenberg model on the square lattice with FM first-neighbor interactions $J_1$ and AF second-neighbor interactions $J_2$~\cite{Sapkota2017,Li2019_Ca}. Close to the value  $\eta = J_1/(2 J_2) = -1$, where the ground state of the classical model transitions from FM order ($\eta < -1$) to stripe-AF
order ($\eta > -1$), the local-moment description captures several features of the INS results. For example, the unusually steep and ridge-like magnetic fluctuation dispersion observed for CaCo$_{2-y}$As$_2$, which is atypical for an A-type antiferromagnet, and the anisotropic shape of the INS peaks at the stripe wavevectors $\mathbf{Q}_\pi=(\pi,0)$ and $(0, \pi)$ are recovered in the local-moment model close to maximal frustration $\eta \approx -1$.

In SrCo$_2$As$_2$, which is the focus of this work, one experimentally extracts a more modest frustration ratio between $\eta \approx -0.5$ at $T=5$~K and $\eta \approx -0.7$ at $T \gtrsim 100$~K from the anisotropy of the INS peaks at $\mathbf{Q}_\pi$~\cite{Li2019_Sr}. This still corresponds to substantial magnetic frustration and one expects stripe-AF fluctuations to prevail at low temperatures. Indeed, INS results indicate that stripe-AF fluctuations develop and suppress FM fluctuations at temperatures below $T \approx 100$~K~\cite{Li2019_Sr}. 
A detailed comparison reveals, however, that one cannot obtain a fully consistent description within a local-moment model. Experimentally, it appears that SrCo$_2$As$_2$ is much more frustrated than expected for $-0.7 < \eta < -0.5$, where stripe-AF fluctuations should clearly dominate over FM ones. Instead, the observed competition between FM and AF fluctuations is much more severe and some properties, such as the size and temperature scale of the FM fluctuations compared to the characteristic magnetic energy scale, can only be captured by a maximally frustrated local moment model with $\eta \approx -1$. This was traced back to a large magnetic energy scale, which has been argued to be more characteristic of itinerant magnets, and led to the characterization of SrCo$_2$As$_2$ as an \emph{itinerant frustrated magnet}~\cite{Li2019_Sr}. 

The notion of itinerant frustration is supported by the observation that SrCo$_2$As$_2$ remains paramagnetic down to the lowest temperatures. The absence of magnetic order in SrCo$_2$As$_2$ was confirmed in NMR measurements down to $50$~mK~\cite{Li2019_Sr}. While magnetic order is absent under hole doping to KCo$_2$As$_2$~\cite{Pandey-PRB-2022, Campbell_Paglione-PRR-2022}, minute amounts of electron doping induce long-range FM order in the Co layers. This suggests a complex and delicate balance between FM and AF fluctuations. For example, Sr(Co$_{1-x}$Ni$_x$)$_2$As$_2$ exhibits long-range magnetic order for $0.013 < x < 0.25$ with a complex helical magnetic structure, where FM Co layers (with moments in the layer) stack to form an incommensurate helix~\cite{Wilde2019}. We note that a symmetry equivalent incommensurate spin-density wave structure is also consistent with the diffraction results. Electron doping via La substitution in Sr$_{1-x}$La$_x$Co$_2$As$_2$ and 2.5\% Nd substitution in Sr$_{0.975}$Nd$_{0.025}$Co$_2$As$_2$~\cite{Shen-SrCo2As2-InChemCommun-2019} leads to the formation of 3D FM order~\cite{Shen2018}. 

Here, we provide a more complete understanding of the intriguing magnetic behavior of doped $\rm{Sr}\rm{Co}_2\rm{As}_2$ by investigating the magnetic susceptibility of a multiorbital itinerant model of doped $\rm{Sr}\rm{Co}_2\rm{As}_2$. We obtain a realistic tight-binding band structure from density functional theory (DFT) calculations and include electronic interactions via a Hubbard-Hund Hamiltonian. The leading magnetic instability in the PM state is found by computing the static transverse spin susceptibility $\chi(\mathbf{q})$ within the random-phase approximation (RPA). This method was previously successfully applied to study the low-temperature phase diagram of multiorbital models of Fe pnictides~\cite{Kuroki2008, Graser2009, Kemper2010, Christensen2017, Christensen2018}.

We map out the weak-coupling RPA magnetic phase diagram as a function of carrier doping $x$ (which, in our convention, is positive for electron doping and negative for hole doping) and the ratio $J/U$ of Hund's coupling $J$ to the Hubbard interaction $U$. Since the ratio $J/U$ that describes the experimental systems is not known exactly, we consider a range of realistic $J/U$ values. It contains a wide region of FM order around the undoped parent compound, but stripe-AF phases appear for sufficiently large hole doping. The critical Hubbard interaction $U_c$ that triggers the development of long-range magnetic order at a given temperature increases under hole doping, but is reduced for electron doping. Since increasing $U$ has similar effects as decreasing the temperature $T$ (it is known to be qualitatively equivalent in the single band case), the behavior of $U_c$ is a good proxy for the expected behavior of the critical temperature $T_c$~\cite{Christensen2018}. 
We choose to tune $U$ in our calculations and fix the temperature to $T=30$~meV as it requires significantly less computational effort than tuning $T$. We note that the RPA is known to overestimate transition temperatures as it neglects certain types of fluctuations~\cite{Christensen2017,Christensen-2016}, and the simulation temperature should thus not be directly compared with the experimental transition temperatures. Rather, our choice of $T=30$~meV arises from balancing the computational demand with the ability to properly resolve spectral features in the bandstructure and the density of states.

Our findings of $U_c(x)$ are thus in good agreement with the experimental observation that magnetic order only occurs under electron doping. Interestingly, we find that $U_c$ at $x=0$ is located close to a shallow minimum, which explains why SrCo$_2$As$_2$ lies on the verge to magnetic ordering and shows a high sensitivity to small changes of electron density - as observed for Ni-doped SrCo$_2$As$_2$.
It can also be related to a peak in the density of states (DOS) that occurs at small positive $x$. We generally associate the theoretically observed slow variation of $U_c$ with $x$ with the phenomenon of itinerant frustration, since it corresponds to an accidental fine tuning of the system close to an instability. Magnetic ordering can thus be induced by small increases in carrier density $x$. Alternatively, we find that $U_c$ decreases for increasing interaction parameter ratio $J/U$, suggesting that systems with larger $J/U$ are more likely to exhibit magnetic order. We note that recent first-principles studies have reported that $J/U$ can be controlled to some extent via pressure or strain~\cite{Panda_Biermann-Pressure_tuning-PRB-2017,Kim-Strain_tuning-PRB-2018}. Pressure tuning of SrCo$_2$As$_2$, however, is complicated by the presence of a structural phase transition to a collapsed tetragonal phase that occurs under pressure~\cite{Jayasekara-PRB-2015}. 

We also quantify the degree of competition between FM and AF fluctuations and identify the leading orbital contributions to the different magnetic states. This allows us to make a direct connection between the orbitally-resolved DOS at the Fermi energy and the leading magnetic instability. Close to an instability towards FM order, the susceptibility is dominated by contributions arising from the Co $d_{xy}$ orbitals, while hybridization of $d_{xy}$ with the other Co $d$ orbitals is much more prominent close to a stripe-AF instability. The closeness of the Fermi energy to the DOS peak arising from partially flat bands with $d_{xy}$ character thus largely determines the type of magnetic instability as a function of doping. While (Stoner) FM order is found if the Fermi level lies close to the DOS peak consisting of $d_{xy}$ (and $d_{z^2}$) states, we observe stripe-AF states when the other orbitals contribute equally to the DOS.
Finally, we calculate the dynamic magnetic susceptibility, $\chi(\mathbf{q}, \omega)$, and find good agreement with INS, demonstrating that the defining features of itinerant frustration are well captured by the interacting multiorbital model. 

The remainder of the article is organized as follows: in Sec.~\ref{sec:model_and_methods} we introduce a realistic 16-band Hubbard-Hund model for SrCo$_2$As$_2$ and describe how to obtain the RPA magnetic susceptibility $\chi(\mathbf{q}, \omega)$. Results for the static susceptibility $\chi(\mathbf{q}, 0)$ as a function of electronic doping $x$, Hubbard $U$, and Hund's coupling $J$ are discussed in Sec.~\ref{sec:susc_results}. We determine the leading magnetic instability and its critical Hubbard $U_c$ as a function of $x$ and $J$. To quantify magnetic frustration we calculate the difference between $U_c$ and the critical $U$ of the first subleading instability. We also discuss the individual orbital contributions to the physical susceptibility, which are markedly different at FM and AF  magnetic instabilities. In Sec.~\ref{sec:dynamics_susc_results}, we present results for the dynamic RPA spin susceptibility $\chi(\mathbf{q}, \omega)$ and in Sec.~\ref{sec:comparison_experiment} we compare theory and experimental results. We present conclusions in Sec.~\ref{sec:Conclusions} and delegate details of the calculations into several Appendices. 

\section{Model and methods}
\label{sec:model_and_methods}
\begin{figure}[t]
    \includegraphics[width=\columnwidth]{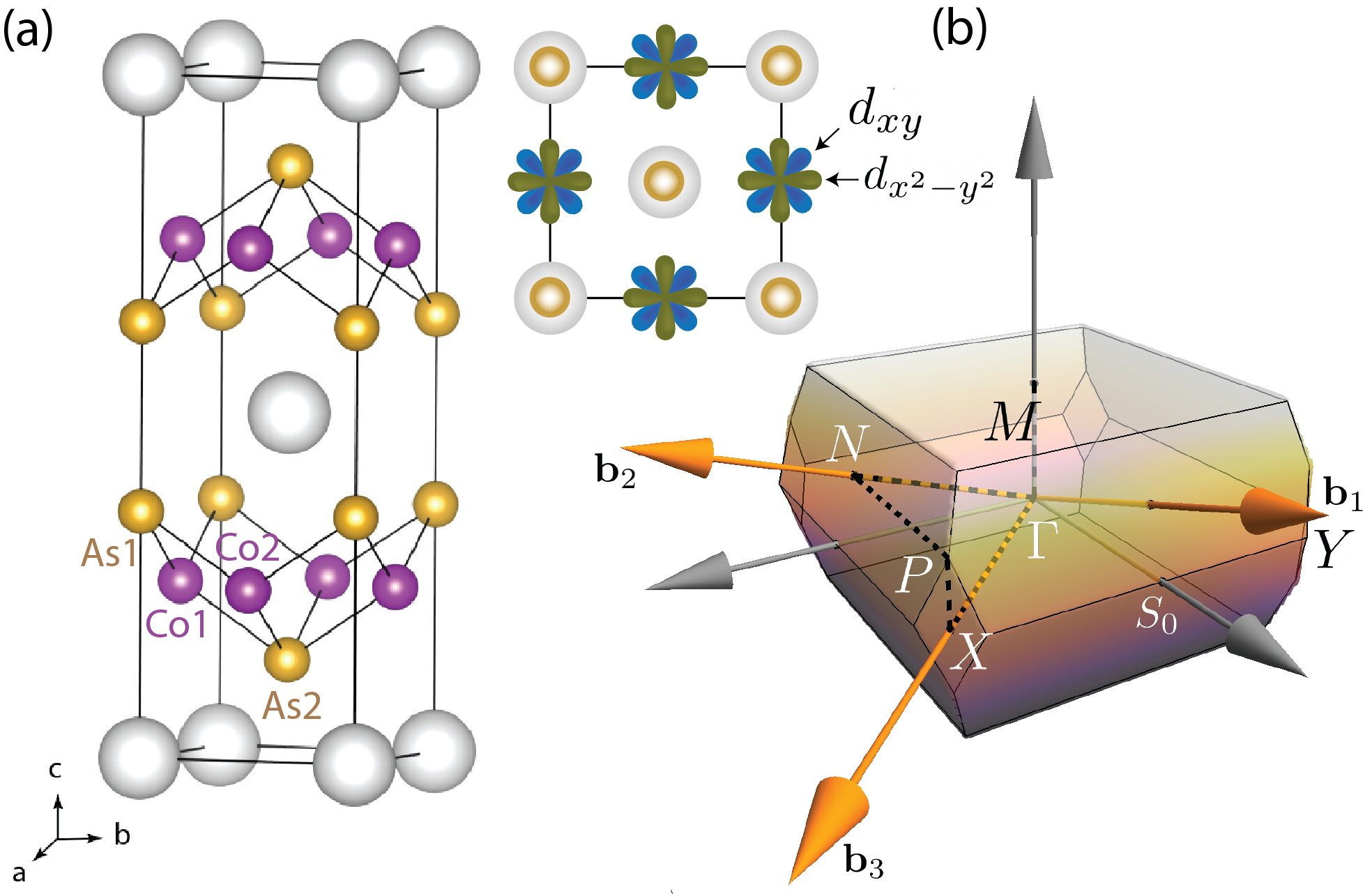}
    \caption{(a) Conventional unit cell of $\rm{Sr}\rm{Co}_2\rm{As}_2$ containing two square layers of Co atoms with As atoms centered above and below. The primitive unit cell contains one Sr, two Co and two As atoms as labelled. The material has a bcc tetragonal crystal structure of ThCr$_2$Si$_2$ type and is described by space group $I4/mmm$ ($c>a$) with $Z=2$ formula units per unit cell. The Sr atoms are located at the Wyckoff site $2a$ $(0 \, 0 \, 0)$, the two equivalent Co atoms are at the $4d$ $(0 \, \frac12 \, \frac14)$ sites and the two equivalent As atoms are at the $4e$ $(0 \, 0 \,z_{\text{As}})$ sites with $z_{\text{As}} = 0.36$. Top view of unit cell depicts $d_{xy}$ (blue) and $d_{x^2 - y^2}$ (olive) orbitals on Co sites. (b) First Brillouin zone with primitive reciprocal lattice vectors $\mathbf{b}_1$, $\mathbf{b}_2$ and $\mathbf{b}_3$ (orange), conventional reciprocal basis vectors (grey vectors) and high-symmetry points. The black dashed line follows the high symmetry path used in Fig.~\ref{fig2:bandstructure+dos}.}
    \label{fig1:unitcell+bz}
\end{figure}

The crystal structure of $\rm{Sr}\rm{Co}_2\rm{As}_2$ lies in the body-centered (bcc) tetragonal symmorphic space group $I4/mmm$ ($\# 139$) with $c>a$ and has the ThCr$_2$Si$_2$ structure type. We focus on the uncollapsed tetragonal structure, where $c \gtrsim 2.8 a$~\cite{Pandey2013,Jayasekara-PRB-2015}. The corresponding point group is $D_{4h}$. The crystal contains square layers of Co atoms with puckered As atoms lying above and below the square centers. The conventional unit cell is shown in Fig.~\ref{fig1:unitcell+bz}(a) and contains two Sr, four Co and four As atoms. The primitive unit cell of the bcc lattice (not shown) contains only one Sr, two Co and two As atoms. A top view of the unit cell shows two Co $d$ orbitals, where $d_{xy}$ has a large spectral weight close to the Fermi energy. In our convention of using a global coordinate system, the Co $d_{xy}$ orbitals point along the nearest-neighbor Co-Co bonds, while the Co $d_{x^2 - y^2}$ orbitals point in between those bonds along the second-neighbor Co-Co bonds. Figure~\ref{fig1:unitcell+bz}(b) depicts the corresponding first Brillouin zone (1BZ) together with the primitive reciprocal lattice vectors (orange) and the conventional ones (grey). In the following, we describe the realistic electronic structure and a downfolded sixteen band Wannier tight-binding model that is valid in a wide region of $\pm 2$~eV around the Fermi energy. We then derive the RPA susceptibility for this model in the presence of electronic interactions.

\renewcommand{\arraystretch}{1.25}
\begin{table}[b]
\centering
\begin{tabular}{*{3}{c}}
\toprule
BZ point & Primitive coordinate & Conventional coordinate\\
\midrule
$X$ & $(0,0,\frac12)$ & $(\frac12, \frac12, 0)$ \\
$P$ & $(\frac{1}{4}, \frac{1}{4}, \frac{1}{4})$ &  $(\frac12, \frac12, \frac12)$\\
$N$ & $(0, \frac12, 0)$ & $(\frac12, 0, \frac12)$\\
$M$ & $(\frac12, \frac12, -\frac12)$ & $(0,0,1)$ \\
\bottomrule
\end{tabular}
\caption{Momentum space coordinates of several high-symmetry points in primitive and conventional unit cell notation.}
\label{tab:high_sym_points}
\end{table}
\renewcommand{\arraystretch}{1}

\subsection{Electronic structure of SrCo$_2$As$_2$}
\label{subsec:electronic_structure}
We obtain a realistic electronic band structure of $\rm{Sr}\rm{Co}_2\rm{As}_2$ using first-principles DFT calculations. 
We neglect the effects of spin-orbit coupling in the DFT calculations, since they are expected to be rather small in the material, since the spin-orbit coupling energy scale of Co is about $70$~meV, but Hubbard and Hund coupling energies are expected to be larger than $1.0$~eV and $0.1$~eV, respectively~\cite{Georges-AnnuRevCondMat-2013}. The details of the DFT approach are discussed in Appendix~\ref{appA:DFT}. We construct a multiorbital tight-binding model from a set of maximally localized Wannier functions (MLWFs) on both the Co- and the As-atoms that are computed by the tool Wannier90~\cite{mostofi2014cpc, Vanderbilt2018, Pizzi2020}. 
We keep all five $d$ orbitals on both Co atoms and all three $p$ orbitals on the two As atoms in the unit cell, resulting in a 16 orbital model. We have checked that including Sr $s$ orbitals in the Wannierization has negligible effects on the tight-binding band structure in the region of $\pm 1$~eV around the Fermi energy. We find that the MLWFs closely resemble the Co $d$-orbitals and the As $p$-orbitals, respectively. We thus use notation that identifies the MLWF with the atomic orbital it approximately represents and introduce the following 16-dimensional Wannier orbital basis vector
\begin{equation}
    \mathbf{\phi}_{\textbf{R}}(\textbf{r}) = \Bigl( \mathbf{d}_{\rm Co1,\textbf{R}}(\textbf{r}), \mathbf{d}_{\rm Co2, \textbf{R}}(\textbf{r}), \mathbf{p}_{\rm As1, \textbf{R}}(\textbf{r}),  \mathbf{p}_{\rm As2, \textbf{R}}(\textbf{r}) \Bigr) \,.
\label{eq:WF_basis}
\end{equation}
Here, $\mathbf{R}$ denotes a Bravais lattice site, the vectors $(\mathbf{d}_{\text{Co1}, \mathbf{R}})_j$ and $(\mathbf{d}_{\text{Co2}, \mathbf{R}})_j$ contain all five atomic $d$ orbitals $j \in \{z^2, xz, yz, x^2 -y^2, xy\}$ at the Co sites, and the vectors $(\mathbf{p}_{\text{As1}, \mathbf{R}})_k$ and $(\mathbf{p}_{\text{As2}, \mathbf{R}})_k$ contain all three $p$ orbitals $k \in \{x, y, z\}$ at the As sites. In the following, we use $a,b = 1, \ldots, 16$ to label the orbital basis: $\phi_{\textbf{R}a}(\textbf{r}) = \braket{\mathbf{r}|\phi_{\textbf{R}a}}$. We work with the tight-binding Hamiltonian matrix
\begin{equation}
    h_{ab}(\textbf{R}) = \langle \phi_{\mathbf{0}a} | H | \phi_{\textbf{R}b} \rangle \,,
\end{equation}
which we obtain from downfolding the DFT bandstructure using the tool Wannier90~\cite{mostofi2014cpc,Vanderbilt2018, Pizzi2020}. Here, $H$ refers to the Hamiltonian used in DFT. We choose to selectively localize the Co $d$ orbitals at the Wyckoff sites that the Co atom occupies in the crystal. This was shown to help preserve the point-group symmetry of the resulting tight-binding Hamiltonian, which can otherwise be weakly violated during the maximal-localization procedure~\cite{Wang_Selective_Localized_Wannier-PRB-2014}. Note that we only selectively localize the ten $d$ orbitals, but not the six $p$ orbitals, which ensures a good tight-binding representation of the DFT band structure.  
\begin{figure*}[t!]
    \includegraphics[width=\linewidth]{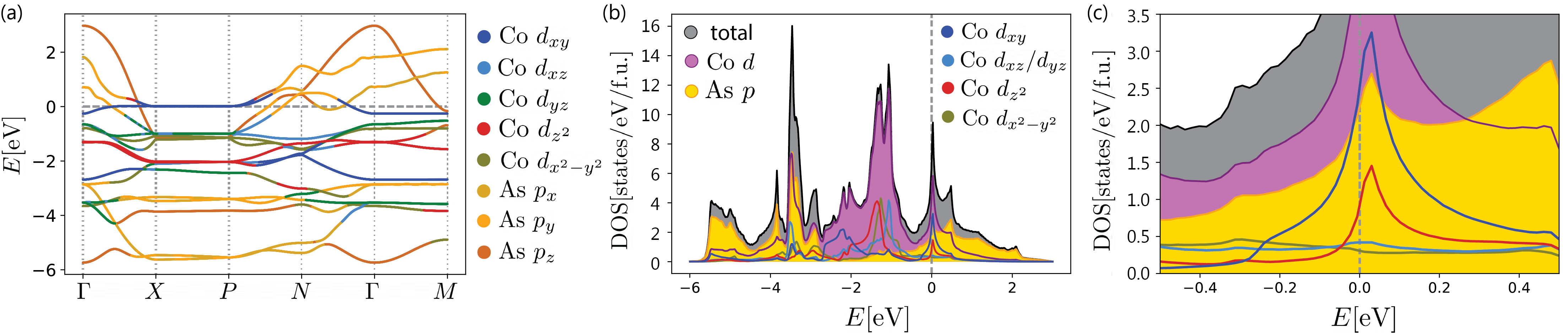}
    \caption{
    (a) Orbitally resolved tight-binding band structure $\varepsilon_n(\mathbf{k})$ for $\rm{Sr}\rm{Co}_2\rm{As}_2$ along a high-symmetry path through the Brillouin zone shown in Fig.~\ref{fig1:unitcell+bz}(b). The bands are colored according to their dominant orbital weight. The Fermi level corresponds to a filling of $n=13$ electrons per spin per unit cell.
    (b) Density of states (DOS) peaks near the Fermi level on the electron-doped side. Different colors denote the orbital contributions to the DOS, which shows that the dominant weight of the peak at $E \approx 0$ is carried by $d$ orbital states of type $d_{xy}$ and $d_{z^2}$. The As weight is about half of the Co weight at the Fermi energy. The orbitals are given in a global coordinate system. (c) Orbitally-resolved DOS near Fermi level.}
    \label{fig2:bandstructure+dos}
\end{figure*}

When going to momentum space, we use the convention to include the orbital basis location $\mathbf{\tau}_a$ (in the primitive lattice vector convention) in the Fourier transform
\begin{equation}
\ket{\phi_{\mathbf{k}a}} = \sum_\mathbf{R} e^{i  \mathbf{k} \cdot (\mathbf{R} + \mathbf{\tau}_a) } \ket{\phi_{\mathbf{R}a}} \,.
\label{eq:fourier_tf_Bloch}
\end{equation}
As shown in detail in Appendix~\ref{appB:symm_conv}, this phase convention that keeps the information of the intra-unit cell placement of the different orbitals through the vector $\mathbf{\tau}_a$ turns out to be crucial to retain the spatial symmetries when calculating the magnetic susceptibility and recover the expected transformation properties under symmetry operations. 
The Hamiltonian matrix elements in the orbital Bloch basis then becomes
\begin{equation}
    h_{ab}(\textbf{k}) = \sum_{\textbf{R}} e^{i\textbf{k}\cdot(\textbf{R} -\mathbf{\tau}_a + \mathbf{\tau}_b)} h_{ab}(\textbf{R}) \,.
    \label{eq:k_transform_def}
\end{equation}
Even with selective localization of the $d$ orbitals at the Co sites, we find that the tight-binding Hamiltonian weakly violates some of the point symmetries of $D_{4h}$. While this is a small effect, we choose to enforce fourfold rotation symmetry: we explicitly symmetrize the Hamiltonian by averaging over points in the Brillouin zone that are related by a four-fold rotation. The procedure is described in detail in Appendix~\ref{appC:Symm_Ham} (a similar procedure is implemented in WannierTools~\cite{Wu2017}). This explicit symmetrization together with the phase convention in Eq.~\eqref{eq:k_transform_def} ensures that the tight-binding model properly obeys the symmetries of the space group $I4/mmm$. Keeping the point-group symmetry of the Wannier eigenstates and energies intact is important to obtain a properly symmetric magnetic susceptibility.

Diagonalization of the tight-binding matrix $h_{ab}(\mathbf{k})$ yields the energy band dispersion $\epsilon_n(\mathbf{k})$, which is shown in Fig.~\ref{fig2:bandstructure+dos}(a) along a high-symmetry path in the 1BZ. A comparison to the full DFT band structure is provided in Fig.~\ref{fig10:comparison_TB_DFT_bs} of Appendix~\ref{appA:DFT}. The appendix also contains the orbitally-resolved Fermi surfaces of the undoped compound in Fig.~\ref{fig12:Fermi_surfaces}. We note again that spin-orbit coupling is neglected and the band structure is thus identical for the two spin states $\sigma = \pm 1$. The Fermi level ($E = 0$) is set to correspond to a filling of $n=13$ electrons per spin per unit cell, corresponding to $6$ electrons per As atom and $7$ electrons per Co atom. Deviations from this filling are parametrized using
\begin{equation}
    x = n - 13 \,,
    \label{eq:def_x_doping}
\end{equation}
where $x > 0$ corresponds to electron and $x < 0$ to hole doping. As shown in Fig. \ref{fig2:bandstructure+dos}(a), the band structure exhibits partially flat bands along the $\Gamma-X$, $\Gamma-M$, and $X-P$ directions. While the first direction describes dispersion arising from electron hoppings within the Co layers, the flatness along $\Gamma-M$, and $X-P$ correspond to weakly dispersing bands along the $k_z$ direction due to a weak coupling between the Co layers. The orbital character of the flat bands is mostly $d_{xy}$, which leads to a pronounced peak in the density of states (DOS) close to (but slightly above) the Fermi energy, as shown in Fig.~\ref{fig2:bandstructure+dos}(b). The plot is obtained for $80\times80\times 80 $ k-points.
Close to the Fermi energy, the DOS is dominated by states with $d_{xy}$ and $d_{z^2}$ orbital weight, while the three other $d$ orbitals and the As $p$-orbitals are subdominant. As noted above, we here use a global coordinate system (or unit-cell coordinate system) when defining the orbitals such that the $d_{xy}$ orbitals point along the nearest-neighbor Co-Co bonds (see Fig.~\ref{fig1:unitcell+bz}). We note that the literature on the Fe pnictides typically uses a local coordinate system, which is rotated by $45$ degrees around the $c$ axis with respect to the global system we use. This rotation results in a permutation of the $d_{xy}$ and $d_{x^2-y^2}$ orbitals.

\subsection{Multiorbital Hubbard-Hund model}
\label{subsec:multiorbital_model}
To study the magnetic spin susceptibility in doped SrCo$_2$As$_2$ we include onsite electronic interactions beyond DFT and consider the following multiorbital Hubbard-Hund Hamiltonian
\begin{equation}
\mathcal{H} = \mathcal{H}_0 + \mathcal{H}_{\rm int}\,.
\end{equation}
Here,
\begin{equation}
    \mathcal{H}_0 = \sum_{\substack{\textbf{k}\sigma \\ ab}}\left(h_{ab}(\textbf{k}) - \mu \delta_{ab} \right)c^{\dagger}_{\textbf{k}a\sigma}c_{\textbf{k}b\sigma}
    \label{eq:DFT}
\end{equation}
is the 16 orbital Wannierized tight-binding model introduced in Sec.~\ref{subsec:electronic_structure}, and the onsite Coulomb interaction part takes the standard Hubbard-Kanamori form~\cite{Gutzwiller1963,Hubbard1963,Kanamori1963,Oles1983}
\begin{widetext}
\begin{equation}
    \mathcal{H}_{\rm int} = U\sum_{\textbf{q} a} n_{\textbf{q}a\uparrow} n_{-\textbf{q}a\downarrow} + \frac{U'}{2} \sum_{\substack{\textbf{q}\sigma\sigma' \\ a\neq b}} n_{\textbf{q}a \sigma} n_{-\textbf{q}b\sigma'}
    +\frac{J}{2} \sum_{\substack{\textbf{k}\textbf{k}'{\textbf{q}} \\ \sigma\sigma' \\ a\neq b}} c_{\textbf{k}+{\textbf{q}} a \sigma}^{\dagger} c_{\textbf{k} b \sigma} c_{\textbf{k}'-{\textbf{q}} b \sigma'}^{\dagger} c_{\textbf{k}' a \sigma'}
    +\frac{J'}{2} \sum_{\substack{\textbf{k}\textbf{k}'\textbf{q} \\ \sigma \\ a\neq b}} c_{\textbf{k}+\textbf{q} a \sigma}^{\dagger} c_{\textbf{k}'-\textbf{q} a \overline{\sigma}}^{\dagger} c_{\textbf{k}' b \overline{\sigma}} c_{\textbf{k} b \sigma}\,.
    \label{eq:Hubbard_int}
\end{equation}
\end{widetext}
The operator $c^{\dagger}_{\textbf{k}a\sigma}$ creates an electron with momentum $\textbf{k}$ in orbital $a \in \{1, \ldots 16\}$ and spin $\sigma \in \{\uparrow, \downarrow\}$; note also that $\overline{\sigma}$ takes the opposite value of $\sigma$. The orbital basis is defined in Eq.~\eqref{eq:WF_basis}. The orbital and momentum dependent density operator is defined as $n_{\textbf{q}a\sigma}=\sum_{\textbf{k}}c^{\dagger}_{\textbf{k}+\textbf{q}a\sigma}c_{\textbf{k}a\sigma}$, where $\mathbf{k}$ runs over the momenta in the 1BZ.
The interaction Hamiltonian contains intra-orbital repulsion with strength $U$ and inter-orbital repulsion with strength $U'$ at the same site. It also contains a Hund's coupling $J$ and a pair-hopping term proportional to $J'$. In the following, we assume spin and orbital rotation invariance. This restricts the parameter space to $J'=J$ and $U' = U - 2J$, such that one is left with two interaction parameters $U$ and $J$.

\subsection{Magnetic spin susceptibility}
\label{subsec:magnetic_susceptibility}
We are interested in calculating both the static and dynamic spin susceptibility in the paramagnetic phase of doped SrCo$_2$As$_2$. The static susceptibility is used to determine the leading magnetic instability as a function of $U$ and $J$, which yields the weak-coupling magnetic phase diagram. Theoretical results for the dynamic susceptibility can be directly compared to INS experimental results and help interpreting these results.

Longitudinal and transverse spin susceptibilities are proportional to each other in the paramagnetic phase and we will thus focus on the transverse part here. The bare transverse spin susceptibility is given by
\begin{equation}
    \chi_{abcd}^{(0)}(\textbf{q}, i\omega_n) = \int_0^{\beta} \mathrm{d}\tau e^{i \omega_n \tau} \left\langle T_\tau S^{+}_{ad}(\textbf{q},\tau) S^{-}_{bc}(-\textbf{q},0) \right\rangle_0\,.\label{eq:spin_susc_def}
\end{equation}
Here $T_\tau$ is the time-ordering operator in imaginary time $\tau$ and $S^{+}_{ab}(\textbf{q},\tau) = \sum_{\textbf{k}} c^{\dagger}_{\textbf{k}+\textbf{q}a\uparrow}(\tau)c_{\textbf{k}b\downarrow}(\tau)$ is the spin raising operator. The spin lowering operator is given by $S^{-}_{ab}(\textbf{q},\tau) = \bigl[ S^{+}_{ab}(-\textbf{q},\tau)\bigr]^\dag$. The brackets $\langle \, \cdot\, \rangle_0$ denote the thermal expectation value with respect to the non-interacting Hamiltonian, $\mathcal{H}_0$, and $\beta \equiv 1/k_B T$ is the inverse temperature. Applying Wick's theorem and performing a summation over Matsubara frequencies we find the bare spin susceptibility
\begin{equation}
    \chi^{(0)}_{abcd}(\textbf{q}, i\omega_n) = \frac{1}{N}\sum_{\textbf{k}, mn} \mathcal{M}_{abcd}^{mn}(\textbf{k},\textbf{q}) \frac{n_F(\varepsilon^{m}_{\textbf{k}})-n_F(\varepsilon^{n}_{\textbf{k}+\textbf{q}})}{\varepsilon^{n}_{\textbf{k+q}}-\varepsilon^{m}_{\textbf{k}}-i\omega_n}\,.
    \label{eq:bare}
\end{equation}
Here, $N$ is the number of unit cells, the labels $m,n$ denote energy bands $\varepsilon^{n}_{\textbf{k}}$ of the tight-binding Hamiltonian $\mathcal{H}_0$ and $n_F(\varepsilon^n_\mathbf{k}) = 1/[e^{\varepsilon^n_\mathbf{k}/T} + 1]$ denotes the Fermi-Dirac distribution function at temperature $T$. Note that the energies $\varepsilon^n_\mathbf{k}$ are defined with respect to the chemical potential $\mu$, which is included in $\mathcal{H}_0$ in Eq.~\eqref{eq:DFT}. 
The tensor $\mathcal{M}_{abcd}^{mn}(\textbf{k},\textbf{q})$ contains information about the orbitals via the eigenfunctions of the Bloch tight-binding Hamiltonian when going from orbital to band space:
\begin{equation}
    \mathcal{M}_{abcd}^{mn}(\textbf{k},\textbf{q}) = u^{n}_{a}(\mathbf{k}+ \mathbf{q})^{\ast} u^{m}_{b}(\textbf{k})^{\ast} u^{n}_{c}(\mathbf{k}+ \mathbf{q}) u^{m}_{d}(\textbf{k}) \,.
    \label{eq:tensor_orbitaltransf}
\end{equation}
Here, $u^{n}_a(\textbf{k})$ is the $n$th eigenstate of $h_{ab}(\textbf{k}) - \mu \delta_{ab}$ in Eq.~\eqref{eq:DFT} at momentum $\textbf{k}$ whose eigenenergy is $\varepsilon^{n}_\textbf{k}$.

We include the effect of onsite Coulomb interactions $\mathcal{H}_1$ through RPA. Diagrammatically, this corresponds to summing all ladder diagrams with no crossing interactions~\cite{Graser2009, Kovacic-PRB-2015}.
The summation involves only interaction processes connected by opposite spins, and can be exactly carried out to yield the RPA spin susceptibility
\begin{equation}
    \chi_{abcd}(\textbf{q},\omega) = \left( \delta_{ae}\delta_{df} - \chi_{eghf}^{(0)}(\textbf{q}, \omega)U^{ga}_{hd} \right)^{-1} \chi^{(0)}_{ebcf}(\textbf{q}, \omega)  \,.
    \label{eq:RPA_spin_susc}
\end{equation}
Here, we have carried out the analytical continuation $i \omega_n = \omega + i \eta_\omega$ and suppressed the infinitesimal $i\eta_\omega$ for brevity. Moreover, $U^{ga}_{hd}$ is given in terms of the $U$, $U'$, $J$, and $J'$ interaction parameters on the Co and As atoms (see Appendix \ref{ap:int}). Note that this expression reduces to the known Stoner formula in the case of a single orbital. One can derive a physical RPA susceptibility that transforms as a scalar (see Appendix~\ref{appB:symm_conv} for details) via the contraction~\cite{Graser2009,Knolle-PRB-2011}
\begin{equation}
    \chi(\textbf{q}, \omega) = \frac{1}{2}\sum_{ab}\chi_{abba}(\textbf{q}, \omega)\,.
    \label{eq:physical_susc}
\end{equation}
A divergence of the static physical spin susceptibility $\chi(\mathbf{q}) \equiv \chi(\mathbf{q}, \omega = 0)$ with infinitesimal $\eta_\omega$ at a specific wavevector $\textbf{Q}$ indicates a weak-coupling magnetic instability and the condensation of magnetic order with ordering vector $\textbf{Q}$. We confine ourselves to ordering wavevectors along the high-symmetry path in the 1BZ shown in Fig.~\ref{fig1:unitcell+bz}(b). The wavevector $\mathbf{Q}$ of the leading magnetic instability is found by increasing the Hubbard interaction parameter $U$, keeping the ratio $J/U$ and the temperature $k_B T = 30$~meV fixed, and recording the first $\textbf{Q}$ among 60 uniformly chosen points along the high-symmetry path for which the physical susceptibility diverges, i.e., 
\begin{equation}
\chi^{-1}(\mathbf{Q}, \omega = 0; U_c, J/U_c, T, \mu) = 0 \,.
\label{eq:phys_susc_diverge}
\end{equation}
Here, we have explicitly added the dependence on $U, J$, $T$ and $\mu$. To map out the behavior under electronic doping, we vary the electronic density per spin per unit cell, $n = 13 + x$, through a rigid shift of the chemical potential $\mu(x,T)$. The dependence of $\mu$ as a function of $x$ is shown in Fig.~\ref{fig11:mu_vs_x} of Appendix~\ref{appA:DFT}. For the summation over $\mathbf{k}$ in Eq.~\eqref{eq:bare} we use a momentum space grid with $25\times 25 \times 25$ $\textbf{k}$-points and we set $\eta_\omega = 3$~meV when computing the static susceptibility.

Finally, we make two remarks. First, our analysis does not determine whether the resulting magnetic order contains only $\textbf{Q}$ or also symmetry related (inequivalent) partners of $\textbf{Q}$. To address this question, one would have to perform a self-consistent mean-field calculation~\cite{Lorenzana2008} or calculate higher-order coefficients of the free energy~\cite{Christensen2017}, which we leave for future studies.
Second, the orbital content of $\chi_{abba}$ on right-hand side of Eq.~\eqref{eq:physical_susc} yields additional information about which orbitals contribute most to the diverging physical susceptibility. We analyze the orbital content of the susceptibility for different doping $x$ and in the different magnetic phases in Sec.~\ref{subsec:orbital_content_magnetic_instab}.

\section{Static spin susceptibility results}
\label{sec:susc_results}
In this section, we present results for the static RPA spin susceptibility $\chi_{abcd}(\mathbf{q}, \omega = 0)$ for both undoped and doped SrCo$_2$As$_2$. We discuss the behavior of the static physical RPA susceptibility $\chi(\mathbf{q})$ in momentum space and map out the leading magnetic instability as a function of $x$ and $J/U$. Then we investigate the competition between FM and stripe-AF fluctuations by calculating the closeness of the first subleading magnetic instability. Finally, we analyze the orbital content of $\chi_{abcd}(\mathbf{q})$ close to the different magnetic instabilities and relate it to the density of states.  

\begin{figure}[t!]
    \includegraphics[width=\linewidth]{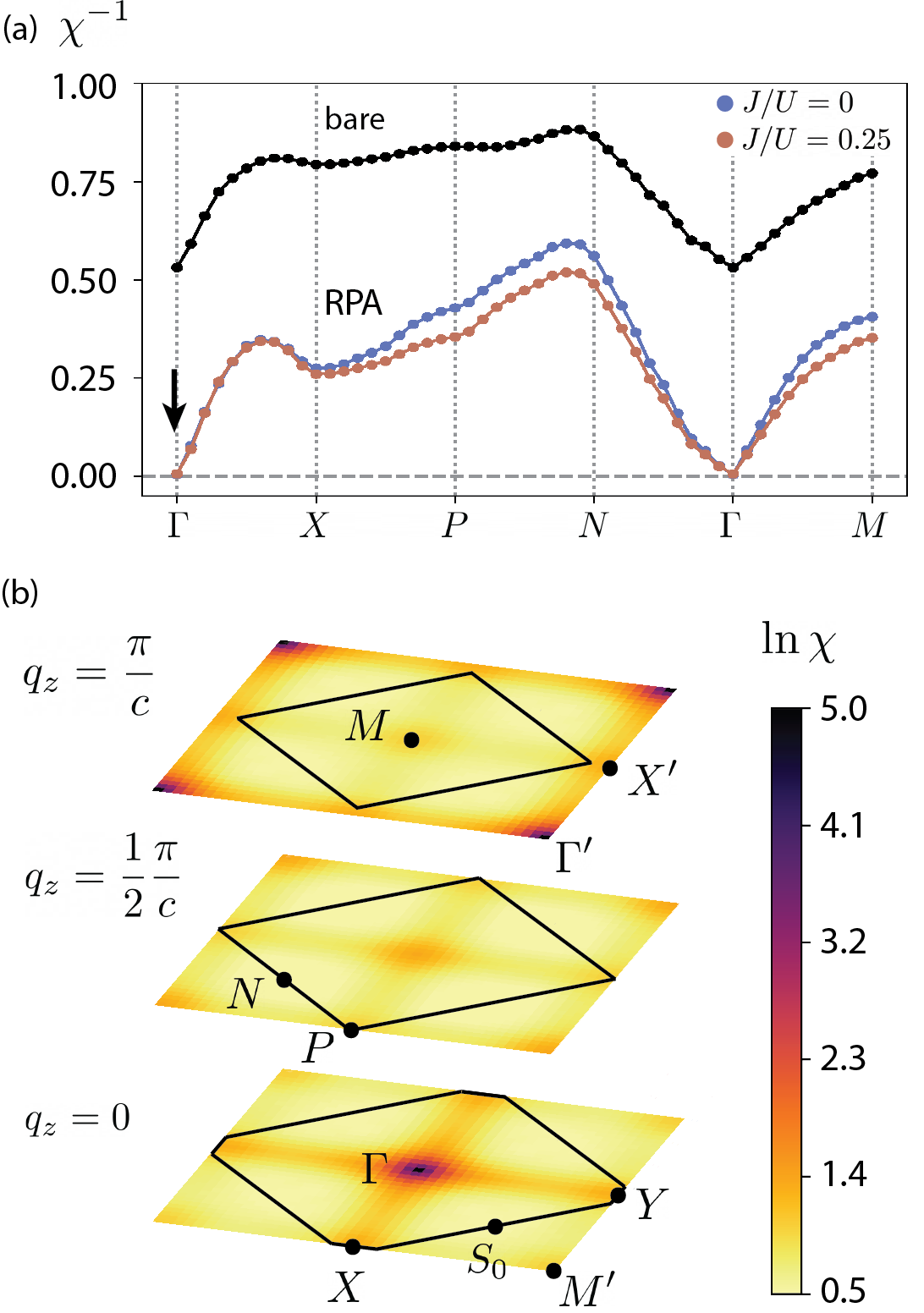}
    \caption{(a) Inverse static bare physical susceptibility $\chi_0^{-1}$ (black) and inverse static RPA physical susceptibility $\chi^{-1}$ (blue, brown) for undoped $\rm{Sr}\rm{Co}_2\rm{As}_2$, plotted along a high-symmetry path in the 1BZ. The plot shows $\chi^{-1}$ for $J/U=0$ (blue) and $J/U=0.25$ (brown) in the vicinity of ordering $U/U_c = 0.995$. The arrow indicates that the leading instability is at $\Gamma$ towards FM order in both cases. (b) Static RPA spin susceptibility $\chi(\mathbf{q})$ as function of $(q_x,q_y)$ for fixed values of $q_z = \{0, \tfrac{1}{2}, 1\} \frac{\pi}{c}$ in conventional coordinates. Interaction parameters are $J/U_c = 0.25$ and $U = 0.995 U_c$. The black line denotes the 1BZ. The temperature and the broadening used in the calculations are $T = 30$~meV and $\eta_\omega = 3$~meV, respectively.}
    \label{fig3:phys_susc_x=0}
\end{figure}

\begin{figure*}[t!]
    \includegraphics[width=\linewidth]{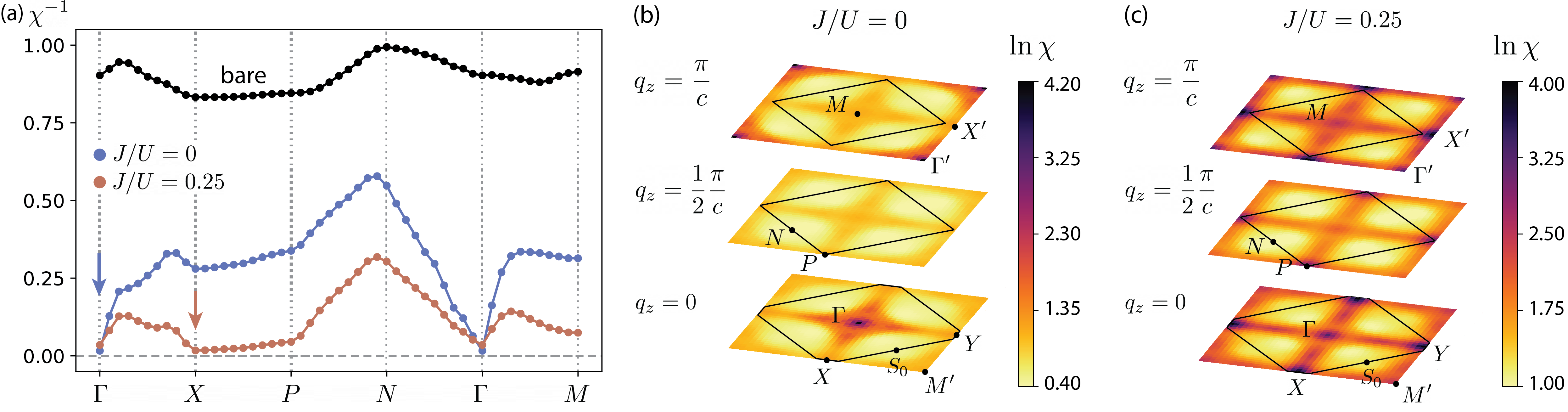}
    \caption{(a) Inverse static bare physical susceptibility $\chi_0^{-1}$ (black) and inverse static RPA physical susceptibility $\chi^{-1}$ (blue, brown) at $x=-0.2$ hole doping, plotted along a high-symmetry path in the 1BZ.
    The plot shows $\chi^{-1}$ for $J/U=0$ (blue) and $J/U=0.25$ (brown) in the vicinity of ordering $U/U_c = 0.995$. In contrast to the undoped case in Fig.~\ref{fig3:phys_susc_x=0}, $\chi$ depends strongly on $J/U$. The arrows indicate the leading instabilities which shift from $\Gamma$ at $J=0$ to $X$ at $J/U_c = 0.25$. Notice the flatness of $\chi^{-1}$ along the $X-P$ direction, which is a sign of itinerant frustration.
	(b, c) Static RPA spin susceptibility $\chi(\mathbf{q})$ as function of $(q_x,q_y)$ for fixed values of $q_z = \{0, \tfrac{1}{2}, 1\} \frac{\pi}{c}$. The interaction parameters are identical to (a) and the black line denotes the 1BZ. Note that the susceptibility exhibits peaks at multiple distinct wavevectors, in particular for stronger Hund's coupling $J/U_c=0.25$. The temperature and the broadening used in the calculations are $T = 30$~meV and $\eta_\omega = 3$~meV, respectively.}
    \label{fig4:phys_susc_x=-0.2}
\end{figure*}

\subsection{RPA susceptibility of the parent compound}
\label{subsec:susc_parent_compound}
In this subsection, we discuss both the bare and RPA spin susceptibility of the undoped parent compound SrCo$_2$As$_2$. Our results show that both the bare and RPA susceptibilities peak at the $\Gamma$ point. The leading instability when increasing interactions is towards FM order with a critical Hubbard $U_c$ that decreases with increasing Hund's coupling $J$.

Figure~\ref{fig3:phys_susc_x=0}(a) shows the inverse physical susceptibility along a high symmetry path in the 1BZ. We observe that the bare susceptibility $\chi_0(\mathbf{q})$ peaks at the $\Gamma$ point and shows a local minimum along the path to the $X$ point. Nonzero interactions enhance this trend and, as a result, the RPA physical susceptibility diverges at $\Gamma$ for all values of $J/U$. The figure shows $\chi^{-1}$ at two different values of $J/U_c$, which qualitatively agree. The critical value of $U$, however, significantly reduces from $U_c(J=0) = 2.10$~eV to $U_c(J/U_c=0.25) = 1.75$~eV as Hund's coupling $J$ increases, which is in agreement with expectations that Hund's coupling favors FM order. 
The fact that the peak in the bare $\chi_0$ determines the ordering vector for sufficiently strong interactions agrees with standard Stoner theory. We will see below that this picture does not always hold true for the doped system. 

Figure~\ref{fig3:phys_susc_x=0}(b) shows a color map of $\chi(\mathbf{q})$ at $U = 0.995 U_c$ and $J/U=0.25$ for three different slices of fixed $q_z$ in the full $2D$ BZ plane. One clearly recognizes the peak at $\Gamma$ from which streaks emerge along the in-plane directions $\Gamma-X$ and $\Gamma-Y$ and to a lesser extent also along the out-of-plane direction $\Gamma-M$. Here, magnetic order at $X = (\frac12, \frac12, 0)$ and $Y = (-\frac12, \frac12,0)$ (in conventional coordinates) corresponds to stripe-AF order. The observed overall behavior of $\chi(\mathbf{q})$ qualitatively remains the same for a wide range of interaction ratios $J/U$ down to $J=0$. We conclude that SrCo$_2$As$_2$ exhibits dominant FM fluctuations and a leading instability towards FM order for all values of $J/U$. This can be traced back to the partially $d_{xy}$-type flat bands along the $X-P$ direction and the resulting large DOS that peaks on the lightly electron doped side (see Fig.~\ref{fig2:bandstructure+dos}).

\subsection{RPA susceptibility of the hole doped system}
\label{subsec:susc_hole_doped}
In this subsection, we present the physical susceptibility of the hole doped system with $x = -0.2$. The behavior of the susceptibility at sufficiently large hole doping $x \lesssim -0.2$ is different from the undoped and lightly doped material.  We find that the bare susceptibility at $x = -0.2$ peaks at the $X$ and $Y$ points, and furthermore, that the leading magnetic instability depends on the interaction ratio $J/U$.

Figure~\ref{fig4:phys_susc_x=-0.2} shows the inverse bare physical susceptibility $\chi_0^{-1}$ at $x = -0.2$ together with the inverse RPA susceptibility $\chi^{-1}$ for two different ratios $J/U_c = 0$ and $J/U_c = 0.25$. The Hubbard interaction is set to be close to the instability $U = 0.995 U_c$, where $U_c(J=0) = 3.16$~eV and $U_c(J/U_c=0.25) = 2.94$~eV. These values are about 50-70\% larger than those at $x=0$, which can be understood from the fact that the DOS is reduced under hole doping (see Fig.~\ref{fig2:bandstructure+dos}). Figure~\ref{fig4:phys_susc_x=-0.2}(a) displays the inverse susceptibilities along a high-symmetry path in the BZ, while Figs.~\ref{fig4:phys_susc_x=-0.2}(b,c) contain $\chi$ as a function of $q_x$ and $q_y$ for three values of $q_z$ and two different Hund's coupling values.
\begin{figure*}[t!]
    \includegraphics[width=\linewidth]{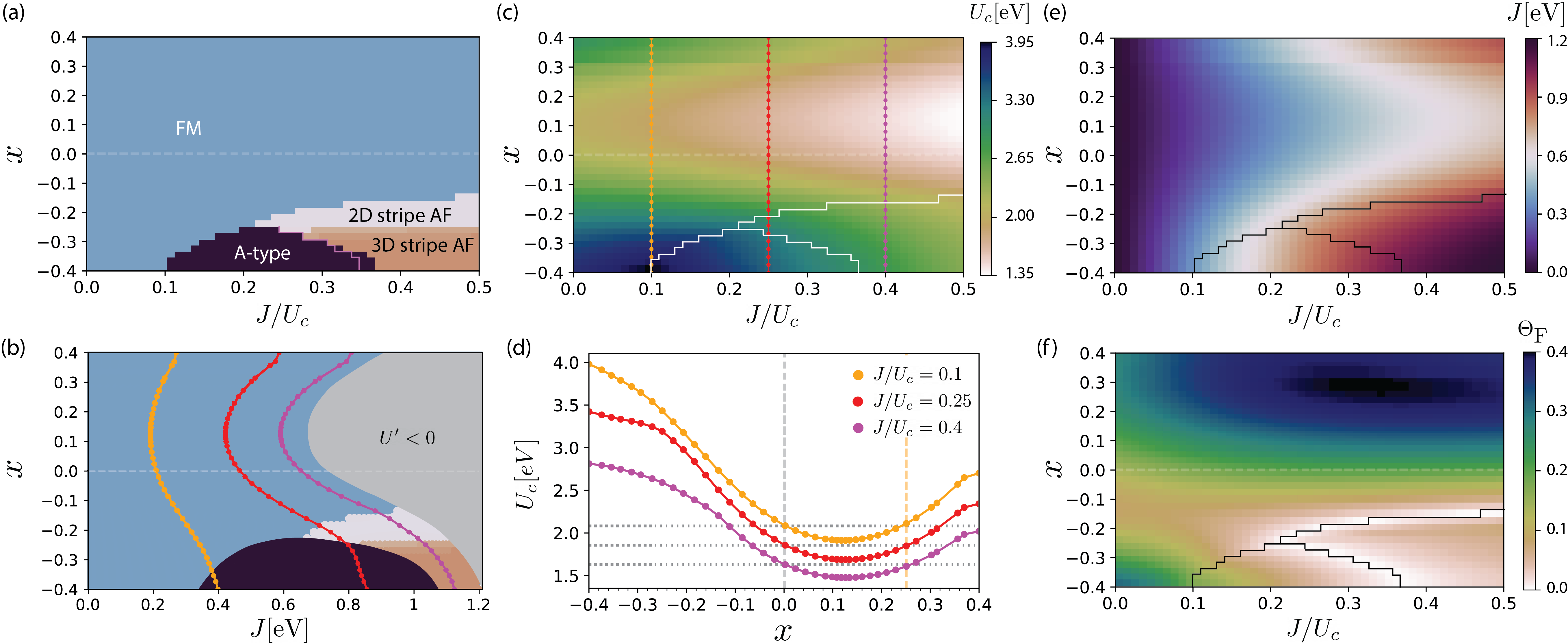}
    \caption{(a) Phase diagram showing the leading magnetic instability as a function of $J/U_c$ and doping $x$. We find four commensurate magnetic orders: FM order with ordering wavevector $\mathbf{Q} = \Gamma=(0,0,0)$, A-type order with $\mathbf{Q} = M=(0,0,1)$, 2D stripe-AF order with $\textbf{Q}=X=(\frac12, \frac12 ,0)$ and 3D stripe-AF order with $\textbf{Q}=P=(\frac12, \frac12, \frac12)$.
    The transition between 2D and 3D stripe-AF occurs through incommensurate phases with ordering vectors $\textbf{Q}=(\frac12, \frac12,\tau)$, where $0 < \tau < \frac12$. The phase diagram is obtained for $U_{\rm{As}}=0$, $J_{\rm{As}}=0$ and the pink lines trace the phase boundaries for $U_{\rm{As}}=U_{\rm{Co}}$ and $J_{\rm{As}} = J_{\rm{Co}}$. (b) Phase diagram with data shown in (a) and smoothed boundaries, but using Hund's coupling $J$ in units of eV as the $x$-axis scale. No data is shown in grey region for which $U'<0$. Colored lines denote cuts at fixed $J/U_c=0.1, 0.25, 0.4$ (yellow, red, purple) [see also panel (c)]. Due to the variation of $U_c$ with $J$, the straight vertical lines in panel (c) appear distorted. 
     (c) Critical Hubbard $U_c$ as a function of $J/U_c$ and $x$. While electron doping ($x>0$) lowers $U_c$, hole doping increases $U_c$. Increasing Hund's coupling $J/U_c$ tends to reduce $U_c$. White lines follow phase boundaries shown in (a). Vertical colored lines denote cuts at fixed $J/U_c$ shown in panel (d). 
     (d) $U_c$ versus $x$ for different values of $J/U_c$ shown in (c). While $U_c$ steeply increases on the hole doped side, $U_c$ is almost flat but slightly decreases on the electron doped side near $x=0$. We associate this behavior with the presence of flat bands (see Fig.~\ref{fig2:bandstructure+dos}). Horizontal dotted lines denote $U_c$ at $x=0$ for the three $J/U_c$ values. The vertical yellow line illustrates the value of $x$ (for $J/U_c=0$) on the electron-doped side which has the same value of $U_c$ as the undoped compound. This bounds the region where we expect magnetic order to exist, since SrCo$_2$As$_2$ is not ordered. The size of this region is almost independent of $J/U_c$.
     (e) The value of $J$ in units of eV at the instability as a function of $J/U_c$ and $x$. Data is obtained directly from $U_c$ shown in panel (c) and the $x$ axis value $J/U_c$.
     (f) Frustration between in-plane ferro- and stripe AF-type phases defined via the parameter $\Theta_{\mbox{F}}$ in Eq.~\eqref{eq:frustration} as a function of interaction ratio $J/U_c$ and filling $x$. Smaller values of $\Theta_F$ correspond to a higher level of frustration. Black lines trace phase diagram of panel (a).
    }
    \label{fig5:PD}
\end{figure*}

Interestingly, while $\chi_0$ peaks at the $X$ point (and the symmetry related $Y$ point, which is not shown), $\chi$ diverges at $\Gamma$ for $J/U = 0$. In contrast, for $J/U=0.25$ the leading instability has shifted to $X$ (and to $Y$). Increasing Hund's coupling thus suppresses FM in favor of stripe-AF order. As we show below in Sec.~\ref{subsec:orbital_content_magnetic_instab}, this can be related to the dominant orbital contributions to $\chi$. While the FM instability is mostly driven by the $xy$ orbital contribution, other orbital components make a larger contribution to stripe-AF. Since Hund's coupling tends to favor alignment of spins in different orbitals, it is generally expected to increase the orbital participation of subleading orbitals $z^2$, $xz$, $yz$ and $x^2-y^2$, which we find to favor stripe-AF over FM. 
Importantly, this is an example where the leading magnetic instability is not determined by the peak in the bare physical susceptibility alone, a multiorbital phenomenon that does not occur for single-band systems with onsite interactions only [see Eq.~\eqref{eq:RPA_spin_susc}].

In addition, we notice that $\chi_0^{-1}$ is nearly-flat and comparable along the $X-P$ and the $\Gamma-M$ directions. It exhibits a local minimum at an incommensurate wavevector along $\Gamma-M$. The flatness along the $k_z$ direction is due to the weak coupling of the Co layers and the resulting flat dispersion along $k_z$. More interesting is that $\chi_{0}$ is of similar size at $\Gamma$ and $X, Y$, which we interpret as a signature of itinerant frustration, as it signals large and comparable fluctuations close to in-plane FM and stripe-AF wavevectors. Close to the instability at $U = 0.995 U_c$, the RPA susceptibility $\chi^{-1}$ still remains flat along the $X-P$ direction, corresponding to a high-degree of competition between in-plane stripe-AF orders with commensurate and incommensurate $q_z$ components. The fact that different magnetic states remain nearly degenerate in the immediate proximity of the magnetic transition is one of the hallmarks of itinerant frustration. The flatness along the $\Gamma-M$ direction, however, is largely lifted close to $U_c$ and there now appears a clear minimum at $\Gamma$ (and a local minimum at $M$ for $J/U_c=0.25$). We conclude that while the parent compound SrCo$_2$As$_2$ is clearly dominated by FM fluctuations for all values of $J/U$, the competition between FM and stripe-AF fluctuations increases with hole doping and is much stronger and dependent on $J/U$ for a system with $x=-0.2$.

\subsection{Phase diagram of leading magnetic instabilities}
\label{subsec:phase_diagram}
\begin{figure*}[t]
    \centering
    \includegraphics[width=\linewidth]{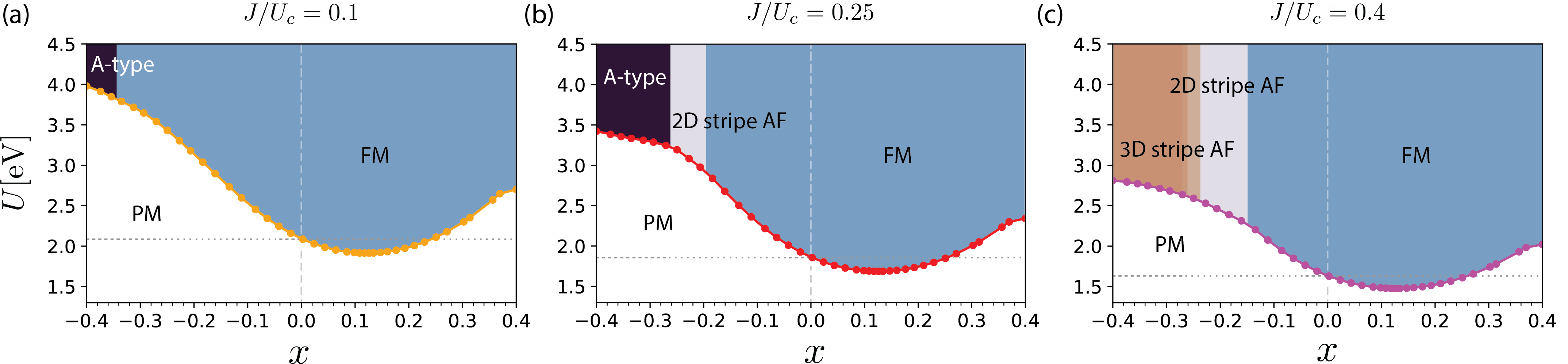}
    \caption{Stoner-type phase diagrams as a function of electronic doping $x$ and Hubbard interaction $U$. Three panels are for Hund's couplings $J/U_c = 0.1, 0.25, 0.4$ from left to right (a-c). The lines denoting $U_c$ are identical to the ones in Fig.~\ref{fig5:PD}(d). We find that for $U < U_c$ the system is paramagnetic (PM) and that the critical Hubbard interaction $U_c$ is minimal for slight electron doping and towards the FM phase. The minimal critical value $\text{min}_x U_c(x)$ at a fixed Hund's coupling $J/U_c$ is reduced from $U_c\approx 2$~eV for $J/U_c=0.1$ to $U_c \approx 1.5$~eV for $J/U_c = 0.4$. As we increase Hund's coupling a stripe-AF phase appears at hole doping $x < -0.2$, whose width increases with $J$. We do not observe any specific feature in $U_c$ in the highly frustrated regime where FM and AF phases meet. }
    \label{fig6:Stoner_PD}
\end{figure*}
This section discusses the phase diagram of the leading magnetic instability as a function of electronic filling $x$ and $J/U$. To obtain the phase diagram shown in Figs.~\ref{fig5:PD}(a,b), we use the same method as described in the context of Figs.~\ref{fig3:phys_susc_x=0} and~\ref{fig4:phys_susc_x=-0.2}. Specifically, for a given filling $x$, we first determine the chemical potential $\mu(x,T=30~\text{meV})$ using the curve shown in Fig.~\ref{fig11:mu_vs_x}. We thus approximate the effect of doping $x$ by a rigid band shift. The covered range in $x \in (-0.4, 0.4)$ corresponds to a rigid shift in the chemical potential $\sim (-0.2,0.1)$~meV. While the approximation of a rigid band shift is justified for small values of $|x|$, it is known that chemical substitution can affect other properties of the system in addition to the electron filling. This includes introducing site disorder and modifications of the lattice structure. Specifically, in Ni-doped SrCo$_2$As$_2$ (i.e. on the electron doped side) the uncollapsed tetragonal phase is found over a wide range of $x < 0.4$ at $T = 300$~K~\cite{Wilde2019}. On the hole doped side, the band structure of KCo$_2$As$_2$ also exhibits features such as a partially flat band between $\Gamma$ and $X$ that are consistent with a rigid band description. 
Given $\mu(x,T)$, we then calculate the bare static susceptibility $\chi^{(0)}_{abcd}(\mathbf{q})$ for $60$ equally spaced points along the high-symmetry path in the BZ shown in Fig.~\ref{fig3:phys_susc_x=0}(a). The chemical potential is set to $\mu(x,T=30~\text{meV})$ and the temperature is $T = 30$~meV. We use a small broadening parameter $\eta_\omega = 3$~meV and sum over $25\times 25 \times 25$ $k$-points in Eq.~\eqref{eq:spin_susc_def}. Computationally, this is the most expensive step in the calculation. Once we have obtained the bare susceptibility, we calculate the physical RPA susceptibility $\chi$ using Eqs.~\eqref{eq:RPA_spin_susc} and~\eqref{eq:physical_susc} for different $U$ at fixed $J/U$. We increase $U$ until $\chi(\mathbf{q})$ diverges, or $\chi^{-1}(\mathbf{q})$ becomes zero at one of the $60$ wavevectors $\mathbf{q}$ along the path in the BZ. 

As shown in Figs.~\ref{fig5:PD}(a,b), we find the leading instability to be towards FM in a wide region around $x=0$, regardless of the value of Hund's coupling $J$. FM prevails for electron doping $x>0$ at all values of $0 \leq J/U_c \leq 0.5$. We note that larger values of $J$ are unphysical as they correspond to inter-orbital attraction due to the relation $U' = U - 2 J$; this region is shown in grey in Fig.~\ref{fig5:PD}(b). On the hole doped side, other magnetic phases appear at $x \lesssim -0.2$ and $J/U_c > 0.1$ (or $J > 0.35 \ {\rm eV}$). In the range of smaller Hund's coupling, a dome of AF A-type magnetic order appears, where FM planes are AF stacked along the $c$ direction. We find the transition from FM to A-type order to be abrupt. At larger Hund's couplings $J/U_c \gtrsim 0.25$ (or $J \gtrsim 0.7$~eV), we observe the emergence of stripe-AF phases, both with $q_z = 0$ and $q_z = \pi/c$. The order with $q_z = 0$ corresponds to the magnetic order observed in the Fe-based 122 systems. The crossover between these two stripe-AF phases occurs gradually via incommensurate magnetic orders with $\textbf{Q}=(\frac12, \frac12,\tau)$, where $0 < \tau < \frac12$.

In Fig.~\ref{fig5:PD}(c), we plot the critical Hubbard $U_c$ required to develop magnetic order versus $x$ and $J/U_c$. One observes that $U_c$ generally decreases with increasing $J$, i.e., Hund's coupling enhances the tendency to develop magnetic order. As a function of $x$, we find that $U_c$ increases quickly under hole doping $x < 0$, as the chemical potential moves away from the peak in the DOS (see Fig.~\ref{fig2:bandstructure+dos}). In contrast, $U_c$ decreases slightly and is almost flat under electron doping as the chemical potential reaches the peak in the DOS. This behavior is independent of the value of $J$ as shown in Fig.~\ref{fig5:PD}(d). We thus conclude that magnetic order is not expected on the hole doped side, but may occur on the electron doped side. The flatness of $U_c$ versus $x >0$ and the fact that one experimentally observes the emergence of magnetic order already for small $x$ in fact suggests that SrCo$_2$As$_2$ lies close to a FM instability. Assuming that $U$ is only slightly below $U_c$ at $x=0$, our calculations predict a region of width $\Delta x \approx 0.2$ where magnetic order should occur on the electron doped side [vertical dashed lines in Fig.~\ref{fig5:PD}(d)]. This is in good agreement with experimental results on Sr(Co$_{1-x}$Ni$_x$)$_2$As$_2$~\cite{Wilde2019}. In Fig.~\ref{fig6:Stoner_PD}, we plot this data in a Stoner-type phase diagram as a function of $x$ and $U$. This phase diagram includes a paramagnetic (PM) phase at $U < U_c$ and the different magnetically ordered phases (FM, A-type, stripe-AF) for $U > U_c$. Our calculations do not show any specific features of $U_c$ across phase boundaries into different magnetic regions, where we expect magnetic frustration to be strongest.

Another finding of our study is that stripe-AF order only emerges at sufficiently large values of the Hund's coupling $J > 0.7$~eV [see Figs.~\ref{fig5:PD}(b,e)]. This may explain the absence of stripe order in real systems, even though the system exhibits prominent stripe-AF fluctuations~\cite{Li2019_Sr}. Finally, we investigate the impact of Hubbard and Hund's interactions on the As sites and find that these play a minor role in the phase diagram. As shown in Fig.~\ref{fig5:PD}(a), the inclusion of significant correlations on the As sites by setting $U_{\text{As}}= U_{\text{Co}}$ and $J_{\text{As}} = J_{\text{Co}}$ shifts the phase boundaries (pink lines) only by a small amount compared to the ones obtained with $U_{\text{As}} = J_{\text{As}} = 0$ (colored phases).

\subsection{Quantifying itinerant magnetic frustration}
\label{subsec:subleading_instabilities}
We can quantify the degree of itinerant magnetic frustration between FM and stripe-AF magnetic order as a function of $x$ and $J/U$ by determining the closeness of the subleading instability. Mathematically, we introduce a frustration parameter that depends on the difference between the critical Hubbard $U_c$ for the leading and the subleading instability
\begin{equation}
    \Theta_{\rm{F}} = \frac{|U_c(\mathbf{Q}_\text{FM}) - U_c(\mathbf{Q}_{\text{AF}})|}{\text{max}\bigl[ U_c(\mathbf{Q}_\text{FM}), U_c^{}(\mathbf{Q}_{\text{AF}})\bigr]}\,.
    \label{eq:frustration}
\end{equation}
Here, $\mathbf{Q}_{\text{FM}} = (0,0,\tau)$ with $0 \leq \tau \leq 1$ is a wavevector that corresponds to in-plane FM order and $U_c(\mathbf{Q}_\text{FM})$ is the minimal Hubbard-$U$ for which $\chi^{-1}(\mathbf{Q}_{\text{FM}}) = 0$. The stripe AF wavevector $\mathbf{Q}_{\text{AF}} = (\frac12,\frac12,\tau)$ with $0 \leq \tau \leq \frac12$  describes the competing stripe-AF order and $U_c(\mathbf{Q}_{\text{AF}})$ is the minimal Hubbard $U$ for which $\chi^{-1}(\mathbf{Q}_{\text{AF}}) = 0$. Small values of $\Theta_F$ thus correspond to high levels of frustration.
We additionally checked that no other ordering vector occurs as a subleading instability along the BZ path in Fig.~\ref{fig1:unitcell+bz}(b), which includes the $\Gamma-X$ and $P-N-\Gamma$ directions. Note that we neglect the presence of magnetic order for $U$ values greater than the critical $U_c$ of the leading instability and are simply increasing $U$ further in our results of the paramagnetic $\chi$ until a competing subleading instability is reached. This still provides a simple approximate method to quantify and compare the degree of frustration in different regions of the phase diagram.

In Fig.~\ref{fig5:PD}(f), we show $\Theta_{\rm{F}}$ as a function of $x$ and $J/U$. First, we observe that, as expected, $\Theta_{\text{F}}$ vanishes at the phase boundaries between FM and AF order. However, we also notice an interesting and nontrivial behavior of $\Theta_\text{F}$: the local minimum of $\Theta_F$ that occurs between A-type and stripe-AF phases at larger hole doping continues into the FM region at smaller doping and reaches $x \approx -0.1$ at $J/U = 0$. The parent compound at $x=0$ is in this sense much more connected to the hole doped region, where $\Theta_F$ is noticeably smaller, than to the electron doped region. At $x=0$ one still finds a substantial amount of itinerant frustration: $\Theta_F(x=0, J/U = 0) = 0.16$.

Comparing Figs.~\ref{fig5:PD}(c) and~(f), we also learn that the behavior of $U_c$ and $\Theta_F$ as a function of $x$ and $J/U$ are quite different. While $U_c$ is correlated with the value of the DOS at the Fermi energy and the size of Hund's coupling $J$, the frustration parameter $\Theta_\text{F}$ is largely determined by the distance to the location of the FM-AF phase boundary. As noted above, it extrapolates the A-type-to-stripe-AF phase boundary line even into the FM regime. This can help explain the puzzling experimental behavior that the magnetic fluctuations in the parent compound ($x=0$) are dominantly stripe-AF at low temperatures (due to the small value of $\Theta_\text{F}$) yet small amounts of electron doping lead to FM order (due to the reduction of $U_c$ by approaching the DOS peak).

\subsection{Orbitally resolved RPA susceptibility}
\label{subsec:orbital_content_magnetic_instab}
In this section, we discuss the orbital resolved contributions $\chi_{abba}$ to the physical RPA susceptibility $\chi = \frac12 \sum_{a,b} \chi_{abba}$. We focus on the behavior close to a magnetic instability and set $U = 0.995 U_c$ in the following. As $U$ approaches $U_c$, the relative weight of the orbital contributions to $\chi_{abba}$ get amplified, but the general trend is already present further away from $U_c$ (we have explicitly checked it at $U = 0.9 U_c$). We find $\chi_{abba}$ to be different for each of the four magnetic instabilities in the phase diagram in Fig.~\ref{fig5:PD}(a). For brevity, we will refer to the four phases by the magnetic ordering wavevector in this section, i.e., refer to FM as $\Gamma$, to A-type as $M$, to 2D stripe-AF as $X$, and to 3D stripe-AF as $P$. We also note that the orbital labeling uses a global (unit cell) coordinate system, shown in Fig.~\ref{fig1:unitcell+bz}. 

\begin{figure}
    \includegraphics[width=\linewidth]{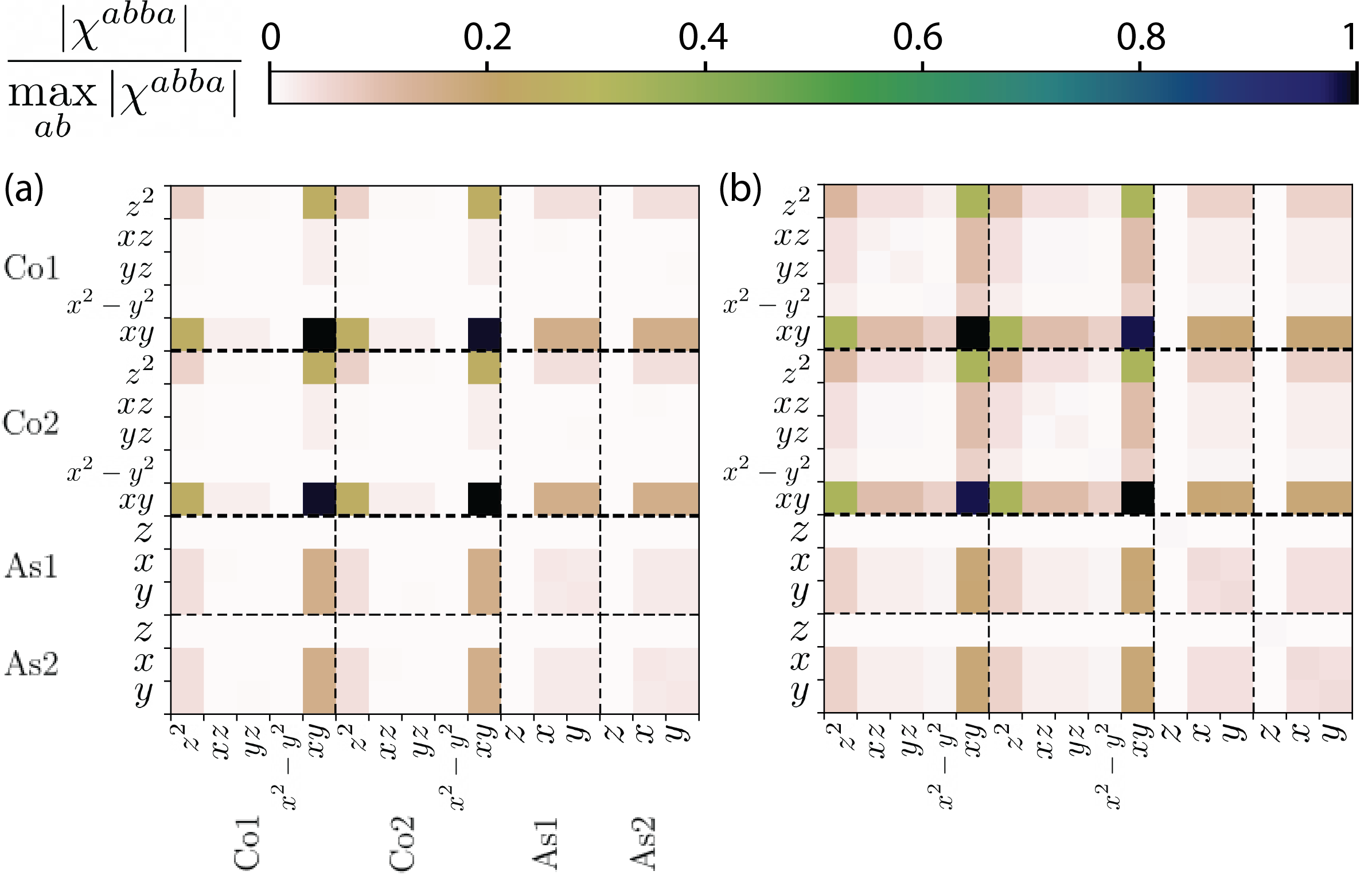}
    \caption{Absolute value of RPA susceptibility elements $\chi_{abba}$ for $x=0$ close to an FM instability. The elements are normalized by the maximal element. Panel (a) is for $J = 0$ and panel (b) is for $J/U_c = 0.4$ at $U = 0.995 U_c$. Here, $T = 30$~meV and $\eta_\omega = 3$~meV.}
    \label{fig7:orb_content_parent_compound}
\end{figure}

\begin{figure}
    \includegraphics[width=\linewidth]{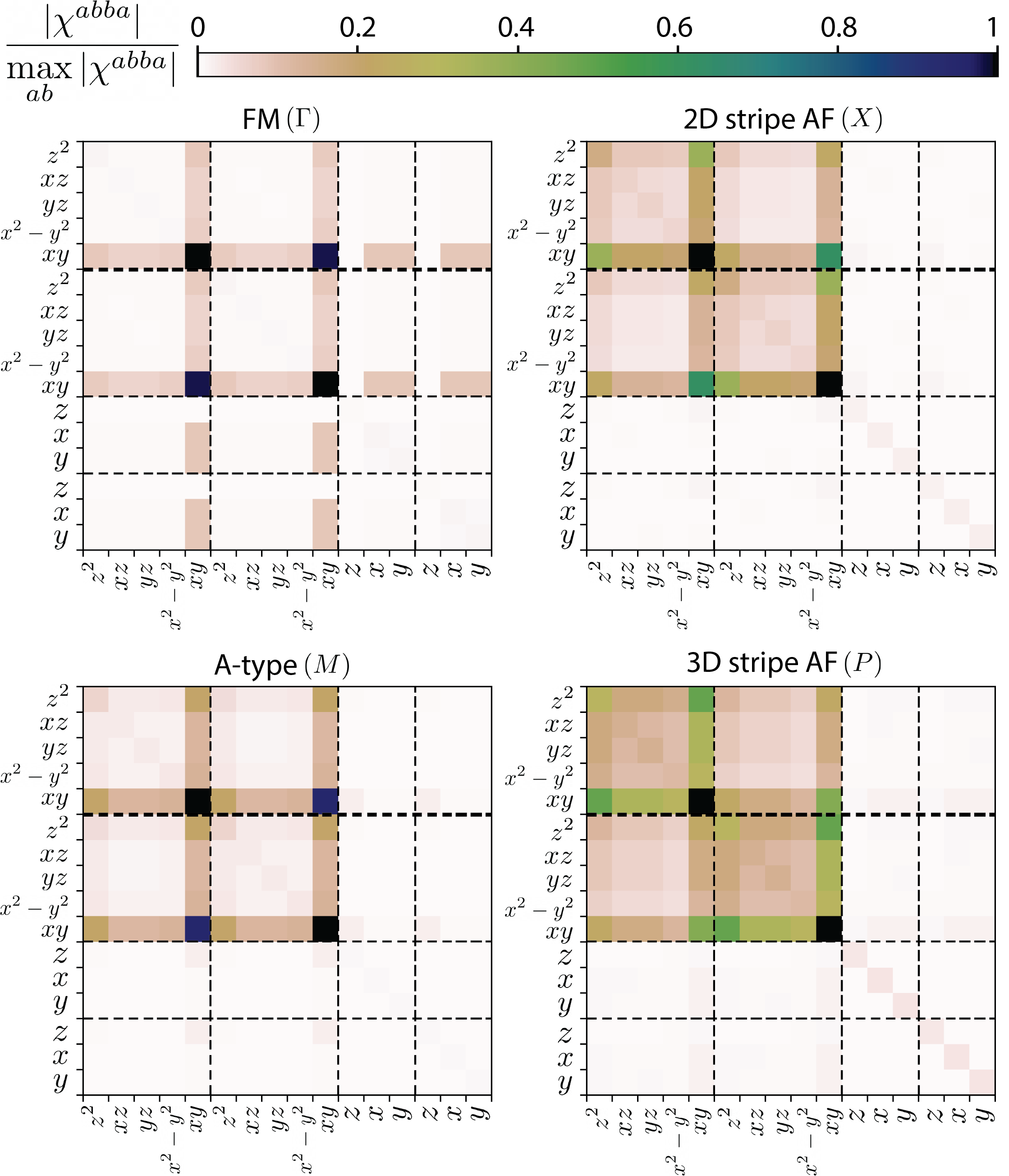}
    \caption{Normalized elements of the RPA susceptibility $\chi_{abba}$ close to four magnetic instabilities, $\Gamma$, $X$, $M$, and $P$, as indicated. First row is for $x = -0.2$ and $J/U_c = 0.2$ (left, $\Gamma$) and $J/U_c = 0.4$ (right, $X$). Second row is for $x = -0.3$ and $J/U_c = 0.2$ (left, $M$) and $J/U_c = 0.4$ (right, $P$). The figures are obtained for $U/U_c=0.995$, $T = 30$~meV, and $\eta_\omega = 3$~meV. The orbital content looks qualitatively the same within each phase at other parameter values, i.e., these results are representative for the four different types of magnetic instabilities in the system. }
    \label{fig8:orb_content_hole_doped}
\end{figure}

The RPA susceptibility components $\chi_{abba}$ that enter the physical susceptibility $\chi = \frac12 \sum_{a,b,} \chi_{abba}$ can be conveniently arranged in a $16 \times 16$ matrix form. In Figs~\ref{fig7:orb_content_parent_compound} and~\ref{fig8:orb_content_hole_doped} we show the absolute value of the components, normalized by the maximum element, as a color matrix plot. Fig.~\ref{fig7:orb_content_parent_compound} contains results for the $x=0$ parent compound at two different values of $J/U = 0, 0.4$, where the leading instability is towards a FM state.
The four different panels in Fig.~\ref{fig8:orb_content_hole_doped} show $\chi_{abba}$ for the four magnetic instabilities $\Gamma$, $M$, $X$, $P$ in the hole doped region at $x = -0.2$ (upper row) and $x = -0.3$ (lower row). While we find some degree of variation of the form of $\chi_{abba}$ within a given phase, the main features are invariant and the four plots are thus representative of the form of $\chi_{abba}$ in the full phase region. 

Let us describe the main features of these plots and the conclusions we draw from it. First, we observe that the diagonal contributions from the $d_{xy}$ orbital are dominant in all four phases, and we find significant contributions from both intrasite elements (Co1-Co1 and Co2-Co2) as well as from the intersite elements (Co1-Co2). The next largest elements at $x=0$ arise from off-diagonal elements between the $d_{xy}$ and $d_{z^2}$ orbitals. As shown in Table~\ref{tab:orb_content}, this can be understood from the large spectral weight of the $d_{xy}$ orbitals at the Fermi energy and the second largest contribution from $d_{z^2}$ at $x =0$. To understand the dominance of $d_{xy}$ elements for $x=-0.3$ we note that our calculations are performed at finite temperature $T = 30$~meV. Even though $d_{xy}$ states at the $x=-0.3$ Fermi energy $\mu = -0.13$~meV have the least weight [see Fig.~\ref{fig2:bandstructure+dos}(a) and Table~\ref{tab:orb_content}], the susceptibility still includes contributions from the region with large $d_{xy}$ weight at finite $T$. Similarly, the DOS peak of the $d_{z^2}$ orbitals at $x=0$ explains that off-diagonal elements between $d_{xy}$ and $d_{z^2}$ are still the largest at $x = -0.3$ and $T > \mu$. 

Second, we notice that increasing Hund's coupling $J$ results in larger off-diagonal elements, since Hund's coupling tends to align spins in different orbitals, i.e., it benefits from electrons occupying and scattering among different orbitals. 
Third, we generally find that close to the stripe-AF instabilities $X$ and $P$, there is a larger hybridization of $d_{xy}$ with the other $d$ orbitals in the susceptibility, including $d_{xz}$, $d_{yz}$, and $d_{x^2-y^2}$. Close to the A-type instability, the susceptibility resembles that of FM with slightly increased $d_{z^2}$ contributions, which may arise from the AF coupling of FM layers along $z$. Finally, we observe that contributions from As $p$ orbitals are negligible, except close to the FM instability, where off-diagonal elements between As $p_x, p_y$ orbitals and $d_{xy}$ contribute about $10\%$ of the relative weight. Note that we here set $U_{\text{As}} = J_{\text{As}} = 0$. This shows that while hybridization with As contributes quantitatively to $\chi$ close to the FM instability, especially at larger $J$, the Co $d$ orbitals are the main factors differentiating between the magnetic instabilities. 

\begin{table}[b]
\centering
\begin{tabular}{*{7}{c}}
\toprule
$x$ & $\mu(0)$[meV] & $\mu(30\text{meV})$ & DOS eV$^{-1}$ & $d_{x^2-y^2}$ & $d_{xz}$, $d_{yz}$ & $d_{z^2}$\\
\midrule
$0$ & $0$ & $-12$ & 7.3 & 0.14 & 0.18 & 0.34\\
$-0.2$ & $-75$ & $-82$ & 2.8 & 0.77 & 0.62 & 0.31\\
$-0.3$ & $-129$ & $-132$ & 2.5 & 2.82 & 2.12 & 1.09\\
\bottomrule
\end{tabular}
\caption{Total DOS per spin at the Fermi energy and different $d$ orbital contributions for different dopings $x$ (see Fig.~\ref{fig2:bandstructure+dos}), chemical potential $\mu(T)$ (in units of meV) at $T=0$ and $T=30$~meV. The DOS contributions of different $d$ orbitals are normalized by the $d_{xy}$ contribution.}
\label{tab:orb_content}
\end{table}

\section{Dynamic RPA spin susceptibility results}
\label{sec:dynamics_susc_results}
In this section, we present results for the imaginary part of the physical RPA susceptibility at finite frequencies $\text{Im} \chi(\mathbf{q}, \omega)$. This allows for a direct comparison with inelastic neutron scattering (INS) results~\cite{Li2019_Sr, Yu_Dai_2019_Sr}. We consider both the parent compound with $x=0$ and a hole doped system with $x = -0.2$ at moderate Hund's coupling $J/U_c = 0.25$. We focus on a Hubbard-$U$ value close to the magnetic instability $U = 0.995 U_c$. The instability is towards FM for $x=0$ and towards 2D-stripe ($X$) at $x = -0.2$. The temperature is set to $T = 30$~meV as in the rest of the paper. 

\begin{figure}
    \includegraphics[width=\linewidth]{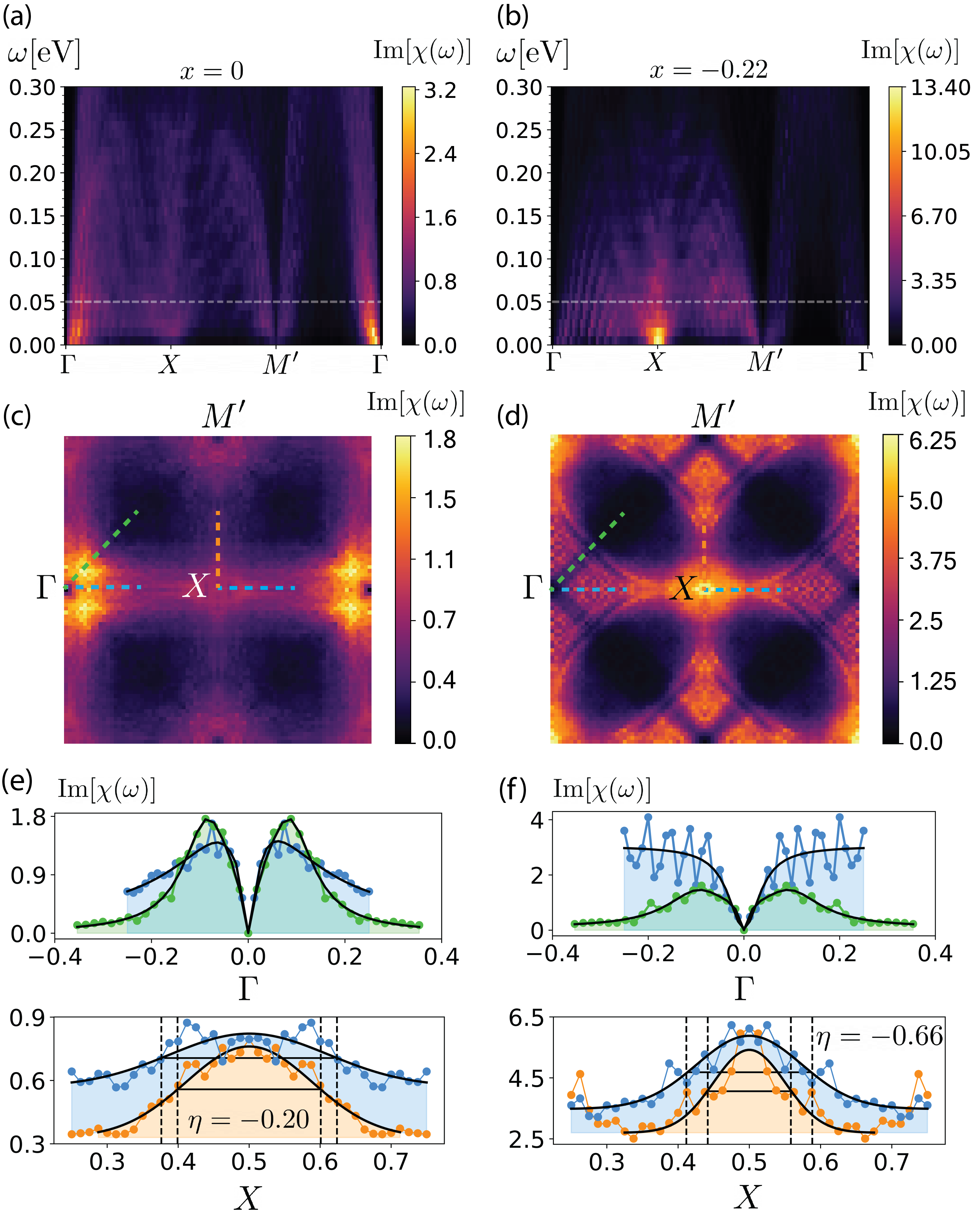}
    \caption{Imaginary part of the dynamical magnetic susceptibility, $\text{Im} \chi(\mathbf{q}, \omega)$, for $x=0$ [panels (a), (c), (e)] and $x = -0.2$ [panels (b), (d), (f)]. Other parameters are $J/U_c = 0.25$, $U = 0.995 U_c$, and $T = 30$~meV. Panels (a,b) show $\text{Im} \chi(\mathbf{q}, \omega)$ along a high-symmetry path in the $q_z = 0$ plane of the BZ. Panels (c,d) show the susceptibility for $(q_x, q_y, q_z = 0)$ at $\omega = 50$~meV [see white dashed line in (a,b)]. Note that the center of the plots corresponds to the $X$ point. Panels (e,f) show 1D cuts close to $\Gamma$ (upper row) and $X$ (lower row) along directions indicated in panels (c,d) with the same color. We extract the frustration parameter $\eta$ from the ratio of the peak widths that we obtain from a Gaussian fit.
    }
    \label{fig9:finite_omega}
\end{figure}

Figure~\ref{fig9:finite_omega} contains results of $\text{Im}\chi(\mathbf{q}, \omega)$ for $x=0$ [panels (a), (c), (e)] and for $x = -0.2$ [panels (b), (d), (f)]. Let us first focus on $x=0$. Fig.~\ref{fig9:finite_omega}(a) shows $\text{Im}\chi(\mathbf{q}, \omega)$ along a high-symmetry path in the $q_z=0$ plane (note that $M'$ is the $M$ point in the second zone, see Fig.~\ref{fig1:unitcell+bz}). We observe well-defined paramagnon modes emerging from the $\Gamma$ point with velocity $v_\Gamma \approx 1100$~meV\;\AA\;along $\Gamma- X$ and a velocity $v_\Gamma \approx 2800$~meV\;\AA\;along $\Gamma- M'$ direction determined from a linear fit in an energy range of $(0-200)$~meV. The width is roughly independent of energy and given by $\gamma_{\Gamma} \lesssim 0.07$\;\AA$^{-1}$. The linewidth is smaller along the $\Gamma-M'$ than along the $\Gamma-X$ direction. Low-energy excitations are also present close to the $X$ and the $M'$ point in the Brillouin zone. The excitations close to $M'$ arise from the small energy scale associated with the weak coupling of the different Co-As layers. In contrast, the presence of low-energy excitations at in-plane momenta $X$ (and $Y$) is a sign of competition and itinerant frustration between FM and stripe-type fluctuations. The large spectral weight around $X$ is also observed experimentally~\cite{Li2019_Sr}.

To investigate the impact of frustration on the dynamic response more systematically, we calculate $\chi(\mathbf{q}, \omega)$ in the complete $q_z = 0$ plane and $\omega = 50$~meV. As shown in Fig.~\ref{fig9:finite_omega}(c) and in agreement with panel (a), we observe a main peak close to $\Gamma$, but also significant spectral weight close to $X$ and $M'$. 
Inspecting the anisotropy of the $\Gamma$ peak, we observe that there is slightly more (about 10\% more) spectral weight along the direction $\Gamma-M'$ than along the direction $\Gamma-X$. The peak around $X$ is also anisotropic and the degree of anisotropy was used in the literature to quantify the degree of itinerant frustration~\cite{Jayasekara2013,Li2019_Sr}. Following these works, we quantify the anisotropy of the elliptical peak at $X$ by the ratio $r$ of the two radii of the ellipse. We relate $r$ to frustration by defining the frustration parameter 
\begin{align}
\eta = \frac{r^2-1}{r^2+1}\,.
\end{align}
The parameter $\eta$ captures the anisotropy of the correlation lengths along two orthogonal directions in the PM phase.
Within a phenomenological local-moment model description of SrCo$_2$As$_2$ using the $J_1$-$J_2$ Heisenberg model on the square lattice, the anisotropy of correlation lengths is related to the ratio of nearest $J_1$ to next-nearest neighbor interactions $J_2$ and one can identify $\eta = J_1/(2 J_2)$~\cite{Jayasekara2013, Li2019_Sr}. Thus, $\eta$ is a direct measure of frustration with $|\eta| = 1$ corresponding to maximal frustration. Here, we obtain $\eta$ from Gaussian fits of the peak in $\chi(\mathbf{q}, \omega)$ close to $X$ and at fixed $\omega = 50$~meV. As shown in Fig.~\ref{fig9:finite_omega}(e), a fit of the spectrum along the $X-M'$ and $X-\Gamma$ directions around the $X$ point yields a moderate anisotropy (or frustration) parameter $\eta = -0.20$ at $x=0$. 
Experimentally, one finds a larger anisotropy in the parent compound with $\eta_{\text{INS}} \approx -0.5$ at low temperatures $T=5$~K~\cite{Jayasekara2013,Li2019_Sr}. The anisotropy increases with temperature and becomes $\eta_{\text{INS}} \approx -0.7$ at $T = 100$~K, where the competition between FM and stripe-AF fluctuations is experimentally most pronounced (FM fluctuations seem to be suppressed below that temperature). In our model, we have to choose parameters closer to the stripe-AF instability in order to reproduce such a large degree of anisotropy. At hole doping $x = -0.2$ and $J/U_c = 0.25$, for example, we find $\eta = -0.66$ [see panel (f) and discussion below]. 

Analyzing the hole doped system at $x = -0.2$ in more detail, we observe in Fig.~\ref{fig9:finite_omega}(b) a steep and broad paramagnon mode emerging from the $X$ point. This agrees with the analysis of the static susceptibility $\chi(\mathbf{q})$ in Fig.~\ref{fig4:phys_susc_x=-0.2}, which diverges at the $X$ point (2D stripe-AF) for $U \rightarrow U_c$. 
The mode has a large stiffness in the range $(0-100)$ meV with a broadening of $\gamma_X \approx 0.07$\;\AA$^{-1}$. We estimate a lower bound on the velocity $v_X \gtrsim 1100$~meV\;\AA \; that is consistent with the lower bound on the transverse velocity of $250$~meV\;\AA \; determined from the INS measurements \cite{Li2019_Sr}.
Fig.~\ref{fig9:finite_omega}(d) contains a two-dimensional map at fixed $\omega = 50$~meV and $q_z = 0$ that shows a main peak at $X$, but also significant spectral weight along the $X-M'$ and $X-\Gamma$ directions. 
As shown in Fig.~\ref{fig9:finite_omega}(f), we observe that the peak amplitude around $\Gamma$ is much smaller along the $\Gamma-M'$ compared to the $\Gamma-X$ direction, which is opposite to our findings at $x=0$. 
The anisotropy of the peak at $X$ is much more pronounced compared to the parent compound and we extract a significant anisotropy (or frustration) parameter of $\eta = -0.66$, as mentioned above.

Regarding the degree of frustration and the anisotropy of the inelastic peak around $X$, we thus conclude from our model calculations that SrCo$_2$As$_2$ behaves experimentally like a slightly hole-doped model that lies closer to the AF instability than the undoped $x=0$ model.

\section{Theory-experiment comparison}
\label{sec:comparison_experiment}
In this section, we compare the results of our model calculations to the experimental ones. First, experimentally one finds SrCo$_2$As$_2$ to be paramagnetic down to the lowest measured temperatures $T = 50$~mK~\cite{Anand2014_1,Li2019_Sr}. This is reproduced in our model when the Hubbard interaction is below the critical value $U < U_c$. Like the experimental system, our model calculations show pronounced anisotropic fluctuations at the $\Gamma$, $X$ and $Y$ points when $U$ lies close to $U_c$. We thus conclude that SrCo$_2$As$_2$ lies on the verge of ordering magnetically with correlations as described by $U$ being only slightly smaller than the required critical $U_c$ value. 

We find that $U_c(x)$ quickly increases under hole doping for all values of $J$, but is almost flat and slightly decreasing under electron doping [see Fig.~\ref{fig5:PD}(d)]. The origin of this behavior is that hole doping moves the chemical potential further away from the peak in the DOS at $\omega \approx 30$~meV, which is caused by a partially flat $d_{xy}$ band that lies just above $E_F$ at $x=0$. Since our calculations are performed at finite temperatures of $T = 30$~meV, the DOS peak is already within the thermal window around $E_F$, and moving $\mu$ closer to the peak by electron doping has not much of an effect. Based on these results, we predict magnetic order to not occur under hole doping, but potentially develop under electron doping, where $U_c$ decreases. This prediction agrees well with experimental observation: while magnetic order has not been found in lightly hole doped materials, already minute amounts of electron doping trigger the development of magnetic order in Sr(Co$_{1-x}$Ni$_x$)$_2$As$_2$ (helical order with in-plane FM) and La$_x$Sr$_{1-x}$Co$_2$As$_2$ (FM). 
We note that the RPA method is known to overestimate transition temperatures since it neglects certain types of fluctuations~\cite{Christensen2017,Christensen-2016}. Thus, the temperature at which the calculations are performed should not be directly compared with experimental transition temperatures. We choose $T=30$~meV to balance the computational costs with a sufficient resolution of spectral features in the band structure and the DOS. The computational complexity increases at lower temperatures since a finer $k$-grid discretization is required to accurately capture sharper features in the susceptibility. The resulting critical $U_c$ that we extract at $T=30$~meV would be reduced if the simulation was performed at a lower temperature, but the phase diagram would not change significantly. 

Our model also captures the essential features associated with itinerant frustration: the presence of both FM and stripe-AF fluctuations and the interesting observation that FM fluctuations dominate only at higher $T > 150$~K~\cite{Li2019_Sr}, but stripe-AF fluctuations take over at lower $T$. To reproduce such a behavior in our model, one needs to consider finite hole doping $x \lesssim -0.1$ and finite Hund's coupling $J/U_c > 0.2$, where we find the leading instability to be at the stripe-AF wavevector [see Fig.~\ref{fig5:PD}(a)]. We generally observe that FM and stripe-AF instabilities lie close to each other in the hole doped region for all values of $J$, even when FM is the leading instability. When we quantify the degree of frustration using $\Theta_F$ in Eq.~\eqref{eq:frustration}, we find the slightly hole doped region ($-0.2 < x < 0$) to be substantially frustrated even at smaller values of $J$, where FM is the leading instability [see Fig.~\ref{fig5:PD}(f)]. We note that tuning $U$ to larger values has a similar effect as lowering temperature and both move the system closer to an instability. We choose to increase $U$ in our calculations as it requires significantly less computational effort than lowering $T$ and it is known (at least in the single band case) to be qualitatively equivalent.

We also find good qualitative agreement between experiment and theory at finite frequencies~\cite{Jayasekara2013,Li2019_Sr,Yu_Dai_2019_Sr}. Theoretically extracting the frustration parameter $\eta$ from the anisotropy of the finite frequency peak of $\chi(\mathbf{q}, \omega)$ at $\mathbf{q} = X$ and $\omega = 50$~meV yields $\eta(x=0,J/U_c=0.25) = -0.20$ and $\eta(x=-0.2,J/U_c=0.25) = -0.66$. This shows that the model at $x= 0$ is slightly less frustrated than the experimental one, for which one finds $\eta_{\text{INS}} \approx -0.5$ at low $T=5$~K and $\eta_{\text{INS}} \approx -0.7$ at higher $T = 100$~K. One can thus reproduce the degree of experimental frustration by moving to the slightly hole doped system $x \approx - 0.2$. Again, this is suggesting that the experimental system SrCo$_2$As$_2$ behaves like a hole doped model and is consistent with conclusions drawn from our static susceptibility results. 

\section{Conclusions}
\label{sec:Conclusions}
In this work, we investigate the magnetic properties of doped SrCo$_2$As$_2$ in a realistic 16-orbital Hubbard-Hund model. By calculating the RPA magnetic susceptibility, we determine the leading magnetic instability as a function of doping $x$ and Hund's coupling $J$. We find FM to be the leading instability in a wide region around $x =0$ and for all values of $J$. Various AF phases that are characterized by the wavevectors $M$, $X$ and $P$ appear under hole doping $x \lesssim -0.1$ and for sufficiently large Hund's coupling $J/U_c > 0.1$. With $U_c\approx 2 - 3$~eV, this corresponds to a Hund's coupling strength $J \approx 200 - 300$~meV, which is realistic~\cite{Georges-AnnuRevCondMat-2013}. We find that the distance of the chemical potential to partially flat bands (with $d_{xy}$ orbital character) and to a resulting DOS peak at $E = 30$~meV determines the value of the critical Hubbard $U$, where magnetic order develops. We observe that $U_c(x)$ experiences a shallow minimum at $x \approx 0.1$ and steeply increases for $x < 0$. This explains the experimental observation that SrCo$_2$As$_2$ develops magnetic order only under electron doping but not under hole doping.

We observe that a larger Hund's coupling $J$ tends to reduce $U_c$ and to distribute electrons among different orbitals, which we find to favor AF phases. We relate this to the orbitally resolved magnetic susceptibility $\chi_{abba}$, which shows significant mixing between the dominant $d_{xy}$ orbital and the other $d$ orbitals only at the AF instabilities. In contrast, $\chi_{abba}$ at the FM instability is dominated by contributions from the $d_{xy}$ orbitals only. This suggests that hole doped systems with a larger $J/U$ could stabilize the sought-after stripe-AF phase. We note that the exact value of $J/U$ in the experimental systems is unknown. Furthermore, recent \emph{ab initio} studies have found that $J/U$ can be tuned over some limited range by applying pressure or strain~~\cite{Panda_Biermann-Pressure_tuning-PRB-2017,Kim-Strain_tuning-PRB-2018}. We leave a detailed theoretical study of possible experimental tuning parameters such as pressure or strain for future work. Note that relatively small pressures trigger a structural transition to a collapsed tetragonal phase, where the DFT electronic band structure is quite different~\cite{Jayasekara-PRB-2015}. 

Our results demonstrate that the phenomenon of itinerant magnetic frustration can be captured within a multiorbital Hubbard-Hund model. In particular, by quantifying frustration as the difference of Hubbard-$U$ values where FM and stripe-AF instabilities occur, $\Theta_F \propto |U_{c,\text{FM}} - U_{c,\text{stripe-AF}}|$, we show that the parent compound experiences a significant degree of frustration. This is signaled by a small value of $\Theta_F$ due to a near degeneracy between FM and stripe-AF orders. Interestingly, we find that the region of small $\Theta_F$ is smoothly connected to the region in the phase diagram at larger hole doping levels, where stripe-AF is the leading instability [see Fig.~\ref{fig5:PD}(f)]. This explains the intriguing experimental fact that the parent compound experiences dominant AF fluctuations at low $T$, but orders FM at minute electron doping. Our study reveals that the first phenomenon is associated with the closeness of the AF instability (i.e. a small value of $\Theta_F$), whereas the second one is due to a reduction of $U_c$ with electron doping. The itinerant magnetic frustration in SrCo$_2$As$_2$ is thus directly tied to the predicted emergence of stripe-AF order at finite hole doping and sufficiently large Hund's coupling $J$. Its paramagnetic behavior is related to the fact that $U < U_c$, i.e., that correlations are slightly too weak to trigger magnetic ordering at $x \leq 0$.

Finally, we calculate the RPA susceptibility $\chi(\mathbf{q}, \omega)$ at finite frequencies and directly relate our findings to inelastic neutron scattering results. We extract the anisotropy $\eta$ of the peaks close to the $X$ point in the Brillouin zone, which was previously related to the degree of frustration of the system. We find that one needs to consider a small amount of hole doping to reproduce the experimental value, since the $x=0$ model is less frustrated than the experimental parent compound SrCo$_2$As$_2$. We generally find that the hole doped model with $x \approx -0.15$ and intermediate values of $J$ best captures the experimental observations on SrCo$_2$As$_2$. Our results offer an alternative and fully itinerant description of the tantalizing phenomenon of itinerant magnetic frustration in doped SrCo$_2$As$_2$, which we find to arise from the interplay of flat band physics and finite Hund's coupling in a correlated multiorbital model. 

We provide all required programs as open-source software, and we make the raw data of our results openly accessible~\cite{Nedic2022}.

\begin{acknowledgments}
We thankfully acknowledge fruitful discussions with David C. Johnston, Milan Kornja\v ca, Andreas Kreisel, Yihua Qiang, Victor L. Quito, and Tha\'{i}s V. Trevisan. The research at Iowa State University and Ames National Laboratory was supported by the U.S. Department of Energy, Office of Basic Energy Sciences, Division of Materials Sciences and Engineering. Ames National Laboratory is operated for the U.S. Department of Energy by Iowa State University under Contract No. DE-AC02-07CH11358.  This computing support for the research reported in this paper in part was supported by the two National Science Foundation grants MRI1726447 and MRI2018594. M.H.C. and R.M.F. were supported by the U.S. DOE, Office of Science, Basic Energy Sciences, Materials Science and Engineering
Division, under award no. DE-SC0020045. The research was funded also in part by the Philip and Virginia Sproul Professorship at Iowa State University. All opinions, findings, and conclusions expressed in this papers are those of the authors.
\end{acknowledgments}

\appendix

\section{Details of the first principle calculations}
\label{appA:DFT}
DFT calculations were carried out using the projected augmented-wave method as implemented in the Vienna \textit{ab initio} simulation package (\textsc{Vasp})~\cite{kresse1999prb,kresse1996cms}.
For the exchange-correlation functional, we employed the Perdew, Burke, and Ernzerhof (PBE)~\cite{perdew1996prl} parametrization in the generalized gradient approximation (GGA).
Experimental lattice parameters ($a=b=3.9471$~\AA, $c=11.801$~\AA, and $z_\text{As}=0.3588$)~\cite{Pandey2013} are used in the calculations, and the plane wave cutoff is set at 300~eV.

\begin{figure}[t]
	\includegraphics[width=\columnwidth]{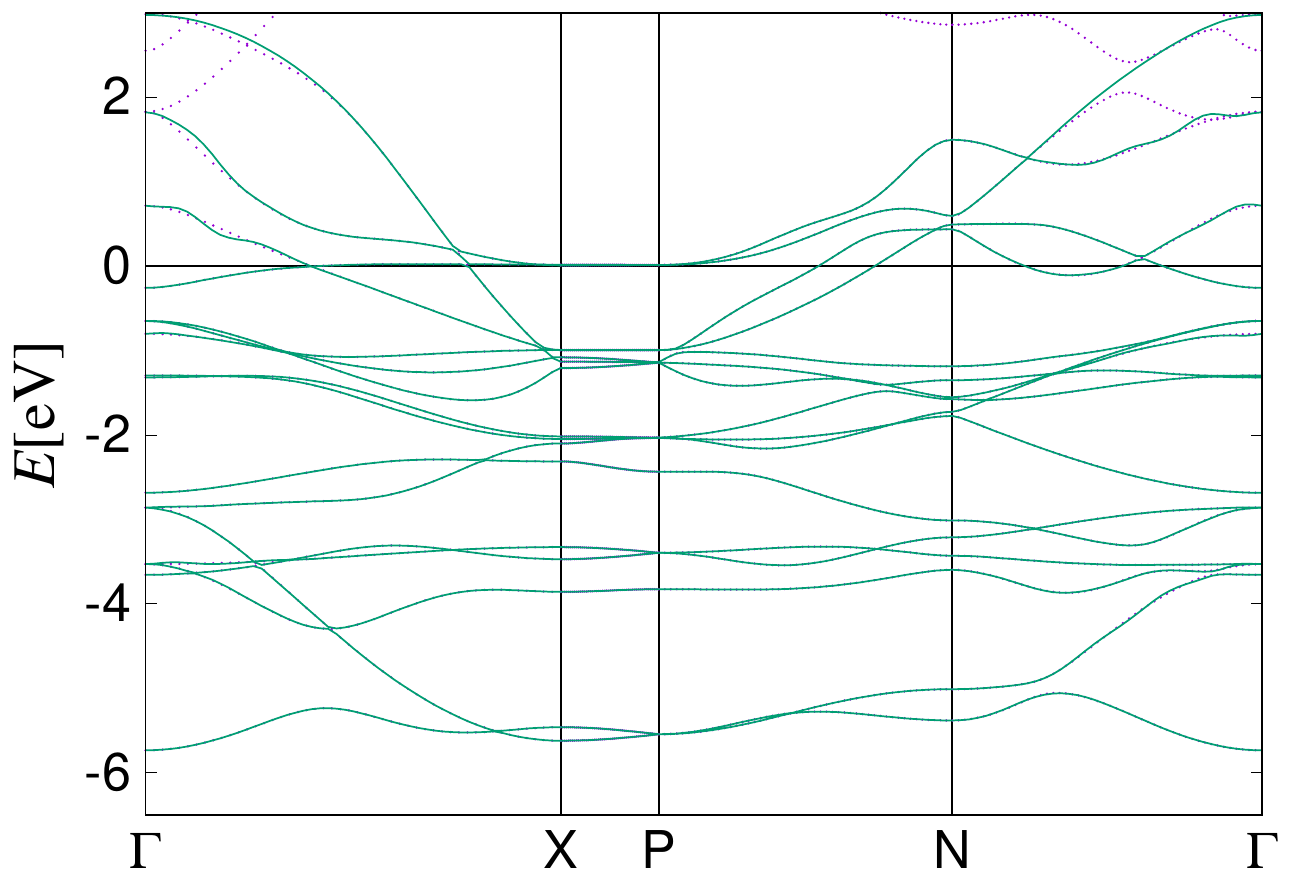}
	\caption{Comparison of the band structure calculated using DFT (dotted line) and the tight-binding model (green lines) along the high symmetry paths as denoted in Fig.~\ref{fig1:unitcell+bz}. }
	\label{fig10:comparison_TB_DFT_bs}
\end{figure}

We constructed the TB Hamiltonian via the maximally localized Wannier functions (MLWFs) method~\cite{marzari1997prb} as implemented in \textsc{wannier90}~\cite{mostofi2014cpc} through a postprocessing procedure~\cite{marzari1997prb,souza2001prb,marzari2012rmp} using the output of the self-consistent DFT calculations.
The basis set consists of 16 MLWFs, corresponding to five Co-$3d$ orbitals and three As-$p$ orbitals on each Co site and As site, respectively.
The selectively localized Wannnier function method was used to ensure the Co-$3d$ orbitals centered on the Co sites.
As shown in Fig.~\ref{fig10:comparison_TB_DFT_bs}, the resulting $16 \times 16$ real-space Hamiltonian $H(R)$ accurately reproduces the band structures in the energy window of interest near the Fermi level, validating its use for susceptibility modeling.
\begin{figure}[t]
    \includegraphics[width=0.95\columnwidth]{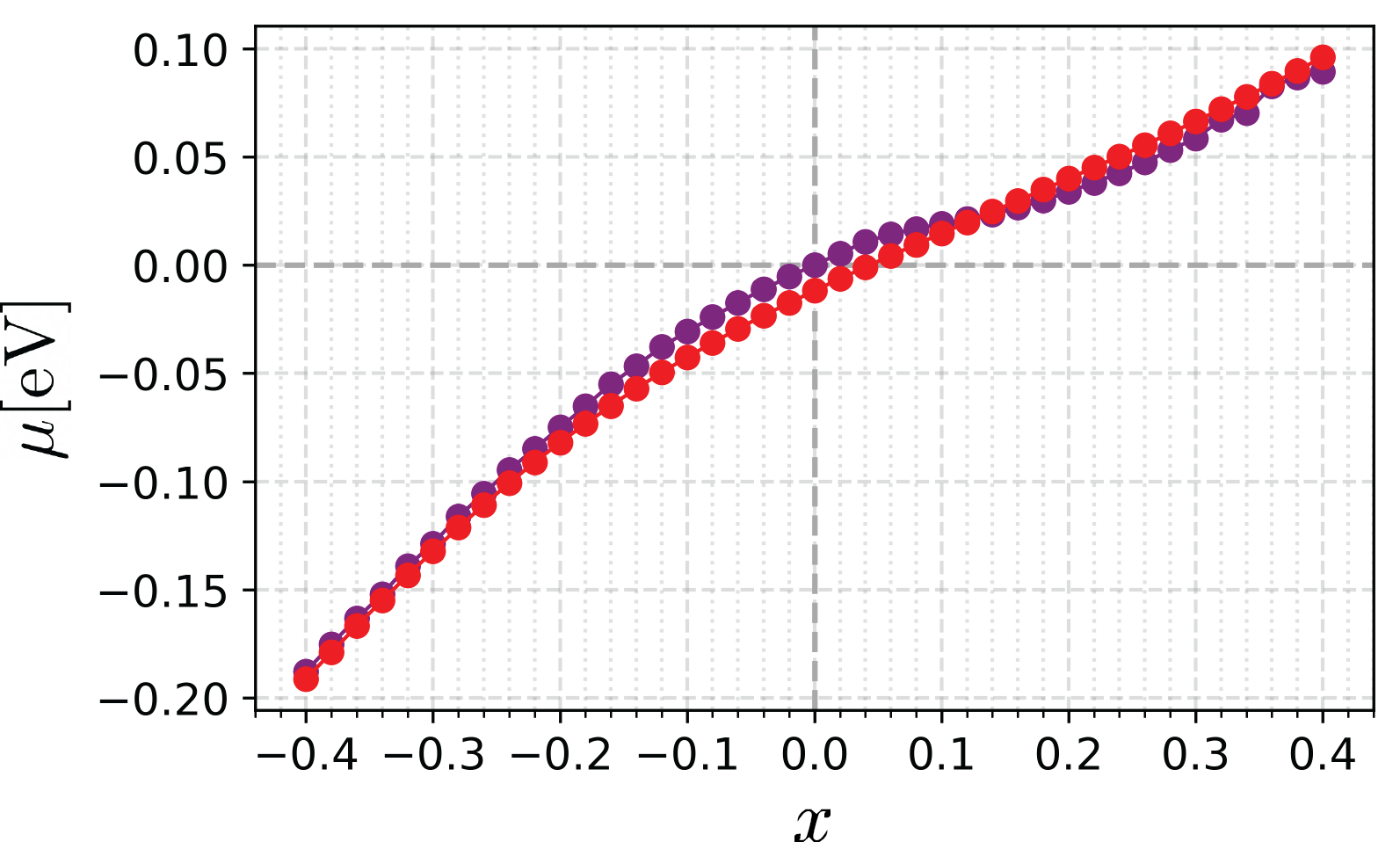}
    \caption{Chemical potential $\mu$ as a function of the electronic density per unit cell $x=n-13$ at $T=0$ (purple) and $T=30$~meV (red).}
    \label{fig11:mu_vs_x}
\end{figure}
The shift in chemical potential as a function of the electronic density per unit cell at $T=0$ and $T=30$~meV is shown in Fig.~\ref{fig11:mu_vs_x}.

\begin{figure}[t]
	\includegraphics[width=\columnwidth]{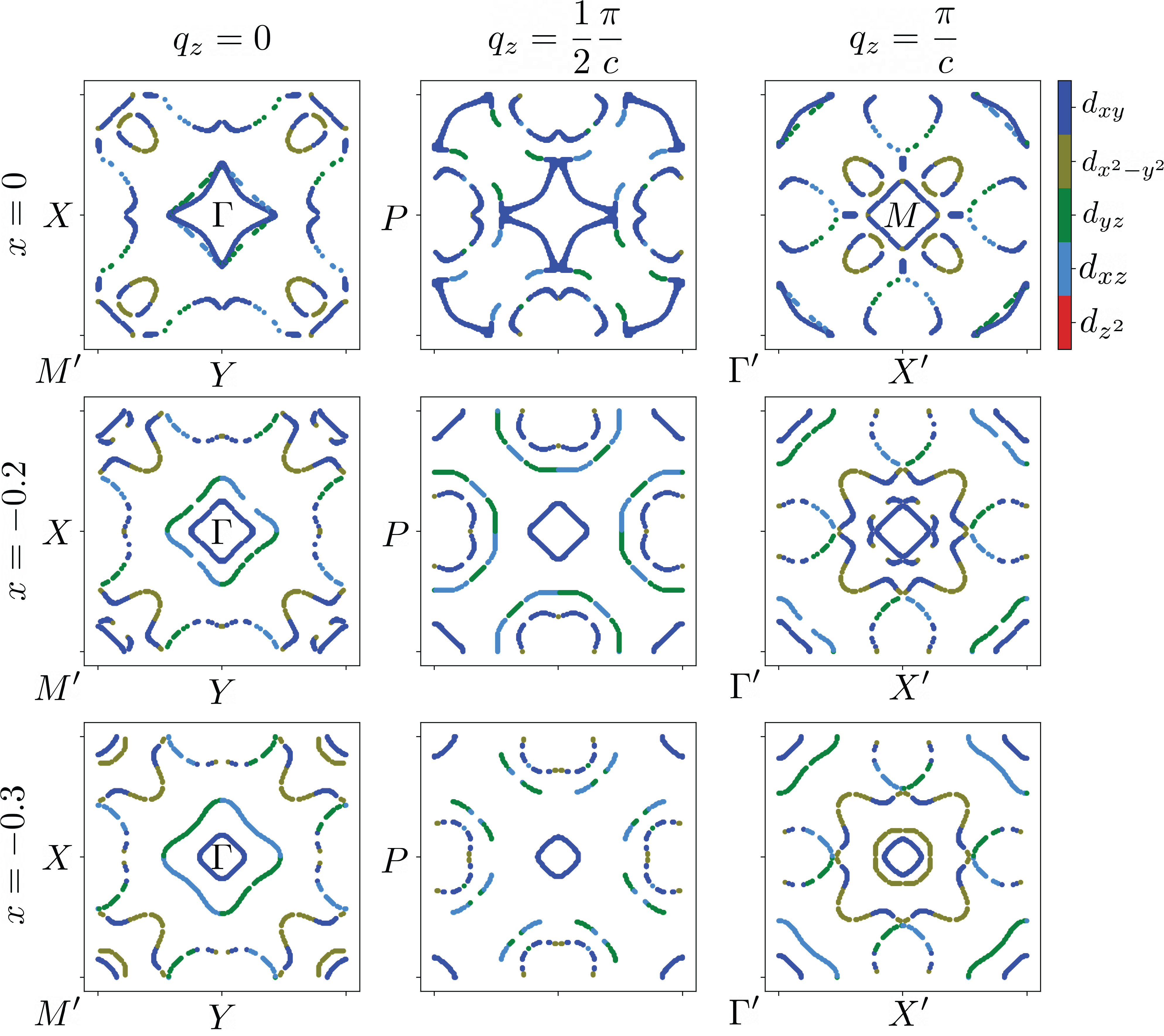}
	\caption{Two-dimensional cuts of the Fermi surface at fixed $q_z = \{0, \tfrac{1}{2}, 1\} \frac{\pi}{c}$ (columns from left to right). Different rows correspond to different dopings: $x=0$ (first row), $x=-0.2$ (second row) and $x=-0.3$ (third row). 
    Color denotes the dominant $d$ orbital weight of the band. Note that some bands have a dominant $p$ orbital character, but we here focus on the $d$ orbital content which is more important for the magnetic instability. We plot bands within an energy window of $\pm 5$~meV around $\omega = 0$ with a discretization of $200\times200$ $k$-points in the 2D cut of the BZ. The figures also include the locations of high-symmetry points.}
	\label{fig12:Fermi_surfaces}
\end{figure}

Fig.~\ref{fig12:Fermi_surfaces} shows two-dimensional cuts of the orbitally-resolved Fermi surfaces for the parent compound $x=0$ and under hole doping $x=-0.2$ and $x=-0.3$. The Fermi surfaces illustrate a quasi-2D character of the system, without signs of a clear nesting. The orbital character of the Fermi surfaces is dominantly $d_{xy}$-like at $x=0$. At finite hole doping, the weight of the $d_{xz}$ and $d_{yz}$ orbitals increases. Hole-doping also promotes the formation of two small electron pockets around the $\Gamma$ and the $M$ point. The dominant orbital character of these pockets is  $d_{xz}, d_{yz}$ for the pocket around $\Gamma$ and is dominantly $d_{x^2-y^2}$ for the pocket around $M$. Notice also the smoothening of the elliptic-like electron pocket around the $X$ and $Y$ points with hole doping resulting in a reduction of the $d_{xy}$ orbital weight in favor or weight from $d_{xz}$ (at $X$) and $/d_{yz}$ orbitals (at $Y$), respectively.

\section{Symmetry transformations for different conventions}
\label{appB:symm_conv}

From the basis of tight-binding orbitals given in Eq.~\eqref{eq:WF_basis}, the basis of Bloch-like wavefunctions can be constructed using one of the conventions \cite{Vanderbilt2018}:
\begin{equation}
\begin{split}
&\psi_a(\textbf{k}) = \sum_{\textbf{R}}e^{i\textbf{k}(\textbf{R}+\tau_a)}\phi_{\textbf{R}a}(\textbf{r})
\\
&\widetilde{\psi}_a(\textbf{k}) = \sum_{\textbf{R}}e^{i\textbf{k}\textbf{R}}\phi_{\textbf{R}a}(\textbf{r})
\end{split}
\label{eq:conventions}
\end{equation}
differing up to a phase factor $e^{i\textbf{k}\tau_a}$ of the fractional positions of atoms within a unit cell. Both conventions give the same band structure, as shown in Fig.~\ref{fig2:bandstructure+dos}(a), while the phase factor difference shows up in the eigenvectors. The two conventions are identical when working with the orbitals of one atom only, e.g. in usual 5-bands models for the Fe-based superconductors.

Using the convention $\psi$, the periodicity in momentum space is not preserved anymore, so one has to work with the Wigner-Seitz cell. With this convention, the eigenvectors at symmetry-related points are related just by a unitary matrix of a point group transformation $\hat{U}$, and non-degenerate eigenvectors differ only up to a global phase $\phi$.
\begin{equation}
u^n_a(\textbf{k}) = \sum_b U_{ab} u^n_b(\mathcal{U}^{-1} \textbf{k})  e^{i\phi}
\label{eq:conv_frac}
\end{equation}

While more commonly used convention $\widetilde{\psi}$ has the advantage of $2\pi$-periodicity of Bloch functions; it has a disadvantage when working with systems with more than one atom per unit cell. When the symmetry operation acting on orbitals centered at one unit cell map them to different unit cells, the transformation matrix between the eigenvectors at the symmetry-related points is a momentum-dependent transformation in orbital space $\hat{\widetilde{U}}(\textbf{k})$.
\begin{equation}
\widetilde{u}^n_a(\textbf{k}) =  \sum_b \widetilde{U}_{ab}(\textbf{k}) \widetilde{u}^n_b(\mathcal{U}^{-1}\textbf{k})  e^{i\phi}
\end{equation}

As an example, the $C_4$ rotation maps orbitals $\rm As_1 \to \rm As_1$, $\rm As_2 \to \rm As_2$, $\rm Co_1 \to \rm Co_2$, $\rm Co_2 \to \rm Co'_1$, where $\rm As_1$, $\rm As_2$, $\rm Co_1$, $\rm Co_2$ all belong to the same unit cell, and $\rm Co'_1$ is positioned in the neighboring unit cell $\textbf{R}'$. Using the convention $\widetilde{\psi}$, the transformation matrix acquires a shift $e^{i\textbf{k} \textbf{R}'}$ for the transformation that includes orbitals outside of the unit cell. As a consequence, the physical susceptibility calculated in Eq.~\eqref{eq:physical_susc} has momentum-dependent transformation under $C_4$ rotation when convention $\widetilde{\psi}$ is used and transforms like a scalar under the symmetries of the system when convention $\psi$ is used.

In what follows, we use convention $\psi$ and prove that the physical susceptibility is invariant under $C_4$ rotations in this convention.

From Eq.~\eqref{eq:bare} and Eq.~\eqref{eq:physical_susc}, the bare physical susceptibility is:
\begin{equation}
    \chi^{(0)}(\textbf{q}) = \frac{1}{2N} \sum_{\substack{ab \\ \textbf{k} \\ mn}} \mathcal{M}_{abba}^{mn}(\textbf{k},\textbf{q}) \frac{n_F(\varepsilon^{m}_{\textbf{k}})-n_F(\varepsilon^{n}_{\textbf{k+q}})}{\varepsilon^{n}_{\textbf{k+q}}-\varepsilon^{m}_{\textbf{k}}}\,.
    \label{eq:bare_abba}
\end{equation}
The eigenvalues for $C_4$ symmetry related points, $\textbf{q}$ and $C_4 \textbf{q}$, are identical and summing over the BZ ($\sum_{C_4 \textbf{k}} = \sum_{\textbf{k}}$), we find
\begin{equation}
    \chi^{(0)}(C_4 \textbf{q}) = \frac{1}{2N} \sum_{\substack{ab \\ \textbf{k} \\ mn}}\mathcal{M}_{abba}^{mn}(C_4 \textbf{k}, C_4\textbf{q}) \frac{n_F(\varepsilon^{m}_{\textbf{k}})-n_F(\varepsilon^{n}_{\textbf{k+q}})}{\varepsilon^{n}_{\textbf{k+q}}-\varepsilon^{m}_{\textbf{k}}}\,,
    \label{eq:bareC4_abba}
\end{equation}
where the tensor $\hat{M}$, introduced in Eq.~\eqref{eq:tensor_orbitaltransf}, contains the information about the transformation of eigenvectors given in Eq.~\eqref{eq:conv_frac} where $\hat{U} = \hat{U}(C_4)$,
\begin{equation}
    \mathcal{M}_{abba}^{mn}(C_4 \textbf{k}, C_4\textbf{q}) = \sum_{a'b'c'd'} U^{T}_{aa'} U^{T}_{bb'} U_{bc'} U_{ad'} \mathcal{M}_{a'b'c'd'}^{mn}(\textbf{k}, \textbf{q})\,.
    \label{eq:bare_phys}
\end{equation}
Note that for physical susceptibility, the acquired global phases cancel exactly.

Finally, using
\begin{equation}
\sum_{a} U^{T}_{aa'} U_{ad'} = \sum_{a} U_{a'a} U_{ad'} =\delta_{a'd'}\,,
\end{equation}
lead to bare physical susceptibilities at $\textbf{q}$ and $C_4\textbf{q}$ being equivalent, while the individual components contributing to $\chi^{(0)}(C_4\textbf{q})$ and $\chi^{(0)}_{\rm{phys}}(\textbf{q})$ will be shuffled.

With Hubbard-like interactions that preserve the symmetry of the original Hamiltonian introduced in Eq.~\eqref{eq:Hubbard_int}, it follows that
\begin{equation}
\chi(\textbf{q}) = \chi(C_4\textbf{q}).
\end{equation}

\section{Symmetrization of the Hamiltonian}
\label{appC:Symm_Ham}

The Wannier Hamiltonian obtained from the first principle calculations is known to slightly break crystal symmetries of the system due to band disentanglement. This symmetry breaking can be controlled for the band structure, enforcing the symmetry conservation within the desired precision at the cost of the agreement between first-principle and fitted band structure. We found this enforcement does not improve the precision of the eigenvectors, which generally differ on third or fourth, but sometimes already on the second digit. Furthermore, using a software package WannierTools to restore the crystal symmetries of the Hamiltonian in real space by generating all rotation matrices and applying them on orbitals on each atom \cite{Wu2017} does not work perfectly on restoring the properties of the eigenvectors. We proceed with restoring the $C_4$ symmetry of our Hamiltonian explicitly, averaging over the symmetry-related points in momentum space.

When sampling the $\textbf{k}$-point mesh in the more complicated Brillouin zones with edges and points shared between more than two neighboring unit cells, it is not trivial to treat boundaries correctly. One often-used approach is the Monkhorst-Pack grid with an even number of points along each direction spanned by primitive lattice vectors that avoid sampling high-symmetry points that usually lie on some boundaries. We deal with this differently by introducing a small constant shift to all points, much smaller than the step size in a sampled grid. The results do not depend on the introduced small shift to the desired precision.

We start by sampling the $\textbf{k}$-point mesh in the reciprocal unit cell, introduce a small constant shift to all points and translate them to the Wigner-Seitz cell. We symmetrize the Hamiltonian in momentum space explicitly to preserve the $C_4$ symmetry of the system, using the convention $\psi$, where $\hat{U} = \hat{U}(C_4)$. The symmetrized Hamiltonian
\begin{widetext}
\begin{equation}
h^{\rm{sym}}_{ab}(\textbf{k}) = \frac{1}{4} \left( h_{ab}(\textbf{k}) + \sum_{cd} {U}_{ac}h_{cd}(C_4^{-1}\textbf{k}){U}_{db}^{-1} + \sum_{cd} {U}_{ac}^2h_{cd}(C_4^{-2}\textbf{k}){U}_{db}^{-2}+ \sum_{cd}{U}_{ac}^3h_{cd}(C_4^{-3}\textbf{k}){U}_{db}^{-3} \right)
\end{equation}
\end{widetext}
conserves the symmetries of the system down to machine precision.

\section{Interaction matrix} \label{ap:int}
\begin{table}[t]
\begin{tabular}{| c c c | c c c| c c c|}
\hline
 \cellcolor{orange!25}$U_{\mbox{As}}$ & $0$ & $0$ & $0$ &  \cellcolor{olive!25}$J_{\mbox{As}}$ & $0$ & $0$ & $0$ &  \cellcolor{olive!25}$J_{\mbox{As}}$ \\ 
 0 & \cellcolor{gray!15}$U'_{\mbox{As}}$ & $0$ & \cellcolor{yellow!45}$J'_{\mbox{As}}$ & $0$ & $0$ & $0$ & $0$ & $0$ \\ 
 0 & $0$ & \cellcolor{gray!15}$U'_{\mbox{As}}$ & $0$ & $0$ & $0$ & \cellcolor{yellow!45}$J'_{\mbox{As}}$ & $0$ & $0$ \\
 \hline
 0 & \cellcolor{yellow!45}$J'_{\mbox{As}}$ & $0$ & \cellcolor{gray!15}$U'_{\mbox{As}}$ & $0$ & $0$ & $0$ & $0$ & $0$ \\ 
  \cellcolor{olive!25}$J_{\mbox{As}}$ & $0$ & $0$ & $0$ & \cellcolor{orange!25}$U_{\mbox{As}}$ & $0$ & $0$ & $0$ &  \cellcolor{olive!25}$J_{\mbox{As}}$ \\ 
 0 & $0$ & $0$ & $0$ & $0$ & \cellcolor{gray!15}$U'_{\mbox{As}}$ & $0$ & \cellcolor{yellow!45}$J'_{\mbox{As}}$ & $0$ \\ 
 \hline
 0 & $0$ & \cellcolor{yellow!45}$J'_{\mbox{As}}$ & $0$ & $0$ & $0$ & \cellcolor{gray!15}$U'_{\mbox{As}}$ & $0$ & $0$ \\  
 0 & $0$ & $0$ & $0$ & $0$ & \cellcolor{yellow!45}$J'_{\mbox{As}}$ & $0$ & \cellcolor{gray!15}$U'_{\mbox{As}}$ & $0$ \\ 
 \cellcolor{olive!25}$J_{\mbox{As}}$ & $0$ & $0$ & $0$ &  \cellcolor{olive!25}$J_{\mbox{As}}$ & $0$ & $0$ & $0$ & \cellcolor{orange!25}$U_{\mbox{As}}$ \\
 \hline
\end{tabular}
\caption{\label{tab:int_matix_As} Onsite interaction on As atom in folded orbital space.}
\end{table}
Here we explicitly write the matrix elements of the interaction of the Hubbard-Hund Hamiltonian introduced in Eq.~\eqref{eq:Hubbard_int} in folded orbital space. The interactions we consider are intra-orbital Coulomb interaction $U^{ll}_{ll}=U$, inter-orbital Coulomb interaction $U^{ll}_{mm}=U^{\prime}$, Hund’s coupling $U^{lm}_{lm} = J$, and the pair-hopping $U^{lm}_{ml} = J^{\prime}$. We distinguish onsite interactions for Co and As atoms. In order to work with matrix equations, we fold indices in the orbital tensor introduced in Eq.~\ref{eq:Hubbard_int} using $A = (ad)$ and $B = (bc)$, such that $U_{dc}^{ab} = U^{AB}$. The interaction matrix for the As (Co) orbitals positioned on a single atom in folded orbital space is a $9 \times 9$ ($25 \times 25$) matrix, given for the As and Co atoms in Table~\ref{tab:int_matix_As} and Table~\ref{tab:int_matix_Co}, respectively.
Rotational invariance in orbital space for each atom separately is satisfied when $J = J'$ and $U'=U-2J$ on each atom.

\begin{table*}[t]
\scalebox{0.9}{
\begin{tabular}{ |c c c c c| c c c c c| c c c c c |c c c c c| c c c c c|}
\hline
\cellcolor{orange!25}$U_{\mbox{Co}}$&0&0&0&0&0&\cellcolor{olive!25}$J_{\mbox{Co}}$&0&0&0&0&0&\cellcolor{olive!25}$J_{\mbox{Co}}$&0&0&0&0&0&\cellcolor{olive!25}$J_{\mbox{Co}}$&0&0&0&0&0&\cellcolor{olive!25}$J_{\mbox{Co}}$ \\
0&\cellcolor{gray!15}$U'_{\mbox{Co}}$&0&0&0&\cellcolor{yellow!45}$J'_{\mbox{Co}}$&0&0&0&0&0&0&0&0&0&0&0&0&0&0&0&0&0&0&0 \\
0&0&\cellcolor{gray!15}$U'_{\mbox{Co}}$&0&0&0&0&0&0&0&\cellcolor{yellow!45}$J'_{\mbox{Co}}$&0&0&0&0&0&0&0&0&0&0&0&0&0&0 \\
0&0&0&\cellcolor{gray!15}$U'_{\mbox{Co}}$&0&0&0&0&0&0&0&0&0&0&0&\cellcolor{yellow!45}$J'_{\mbox{Co}}$&0&0&0&0&0&0&0&0&0 \\
0&0&0&0&\cellcolor{gray!15}$U'_{\mbox{Co}}$&0&0&0&0&0&0&0&0&0&0&0&0&0&0&0&\cellcolor{yellow!45}$J'_{\mbox{Co}}$&0&0&0&0 \\
\hline
0&\cellcolor{yellow!45}$J'_{\mbox{Co}}$&0&0&0&\cellcolor{gray!15}$U'_{\mbox{Co}}$&0&0&0&0&0&0&0&0&0&0&0&0&0&0&0&0&0&0&0 \\
\cellcolor{olive!25}$J_{\mbox{Co}}$&0&0&0&0&0&\cellcolor{orange!25}$U_{\mbox{Co}}$&0&0&0&0&0&\cellcolor{olive!25}$J_{\mbox{Co}}$&0&0&0&0&0&\cellcolor{olive!25}$J_{\mbox{Co}}$&0&0&0&0&0&\cellcolor{olive!25}$J_{\mbox{Co}}$ \\
0&0&0&0&0&0&0&\cellcolor{gray!15}$U'_{\mbox{Co}}$&0&0&0&\cellcolor{yellow!45}$J'_{\mbox{Co}}$&0&0&0&0&0&0&0&0&0&0&0&0&0 \\
0&0&0&0&0&0&0&0&\cellcolor{gray!15}$U'_{\mbox{Co}}$&0&0&0&0&0&0&0&\cellcolor{yellow!45}$J'_{\mbox{Co}}$&0&0&0&0&0&0&0&0 \\
0&0&0&0&0&0&0&0&0&\cellcolor{gray!15}$U'_{\mbox{Co}}$&0&0&0&0&0&0&0&0&0&0&0&\cellcolor{yellow!45}$J'_{\mbox{Co}}$&0&0&0 \\
\hline
0&0&\cellcolor{yellow!45}$J'_{\mbox{Co}}$&0&0&0&0&0&0&0&\cellcolor{gray!15}$U'_{\mbox{Co}}$&0&0&0&0&0&0&0&0&0&0&0&0&0&0 \\
0&0&0&0&0&0&0&\cellcolor{yellow!45}$J'_{\mbox{Co}}$&0&0&0&\cellcolor{gray!15}$U'_{\mbox{Co}}$&0&0&0&0&0&0&0&0&0&0&0&0&0 \\
\cellcolor{olive!25}$J_{\mbox{Co}}$&0&0&0&0&0&\cellcolor{olive!25}$J_{\mbox{Co}}$&0&0&0&0&0&\cellcolor{orange!25}$U_{\mbox{Co}}$&0&0&0&0&0&\cellcolor{olive!25}$J_{\mbox{Co}}$&0&0&0&0&0&\cellcolor{olive!25}$J_{\mbox{Co}}$ \\
0&0&0&0&0&0&0&0&0&0&0&0&0&\cellcolor{gray!15}$U'_{\mbox{Co}}$&0&0&0&\cellcolor{yellow!45}$J'_{\mbox{Co}}$&0&0&0&0&0&0&0 \\
0&0&0&0&0&0&0&0&0&0&0&0&0&0&\cellcolor{gray!15}$U'_{\mbox{Co}}$&0&0&0&0&0&0&0&\cellcolor{yellow!45}$J'_{\mbox{Co}}$&0&0 \\
\hline
0&0&0&\cellcolor{yellow!45}$J'_{\mbox{Co}}$&0&0&0&0&0&0&0&0&0&0&0&\cellcolor{gray!15}$U'_{\mbox{Co}}$&0&0&0&0&0&0&0&0&0 \\
0&0&0&0&0&0&0&0&\cellcolor{yellow!45}$J'_{\mbox{Co}}$&0&0&0&0&0&0&0&\cellcolor{gray!15}$U'_{\mbox{Co}}$&0&0&0&0&0&0&0&0 \\
0&0&0&0&0&0&0&0&0&0&0&0&0&\cellcolor{yellow!45}$J'_{\mbox{Co}}$&0&0&0&\cellcolor{gray!15}$U'_{\mbox{Co}}$&0&0&0&0&0&0&0 \\
\cellcolor{olive!25}$J_{\mbox{Co}}$&0&0&0&0&0&\cellcolor{olive!25}$J_{\mbox{Co}}$&0&0&0&0&0&\cellcolor{olive!25}$J_{\mbox{Co}}$&0&0&0&0&0&\cellcolor{orange!25}$U_{\mbox{Co}}$&0&0&0&0&0&\cellcolor{olive!25}$J_{\mbox{Co}}$ \\
0&0&0&0&0&0&0&0&0&0&0&0&0&0&0&0&0&0&0&\cellcolor{gray!15}$U'_{\mbox{Co}}$&0&0&0&\cellcolor{yellow!45}$J'_{\mbox{Co}}$&0 \\
\hline
0&0&0&0&\cellcolor{yellow!45}$J'_{\mbox{Co}}$&0&0&0&0&0&0&0&0&0&0&0&0&0&0&0&\cellcolor{gray!15}$U'_{\mbox{Co}}$&0&0&0&0 \\
0&0&0&0&0&0&0&0&0&\cellcolor{yellow!45}$J'_{\mbox{Co}}$&0&0&0&0&0&0&0&0&0&0&0&\cellcolor{gray!15}$U'_{\mbox{Co}}$&0&0&0 \\
0&0&0&0&0&0&0&0&0&0&0&0&0&0&\cellcolor{yellow!45}$J'_{\mbox{Co}}$&0&0&0&0&0&0&0&\cellcolor{gray!15}$U'_{\mbox{Co}}$&0&0 \\
0&0&0&0&0&0&0&0&0&0&0&0&0&0&0&0&0&0&0&\cellcolor{yellow!45}$J'_{\mbox{Co}}$&0&0&0&\cellcolor{gray!15}$U'_{\mbox{Co}}$&0 \\
\cellcolor{olive!25}$J_{\mbox{Co}}$&0&0&0&0&0&\cellcolor{olive!25}$J_{\mbox{Co}}$&0&0&0&0&0&\cellcolor{olive!25}$J_{\mbox{Co}}$&0&0&0&0&0&\cellcolor{olive!25}$J_{\mbox{Co}}$&0&0&0&0&0&\cellcolor{orange!25}$U_{\mbox{Co}}$ \\
\hline
\end{tabular}
}
\caption{\label{tab:int_matix_Co} Onsite interactions on Co atom in folded orbital space.}
\end{table*}

The full interaction matrix for $16$ orbitals is a block-diagonal $256 \times 256$ matrix in folded orbital space consisting of 2 Co blocks of size $80 \times 80$ and 2 As blocks of size $48 \times 48$. Each of these blocks are sparse matrices with non-zero elements on the positions of the on-site interactions for Co and As atoms.

\newpage
\bibliographystyle{apsrev4-2}
\bibliography{multiorbitalRPA_SrCo2As2.bib}

\end{document}